\documentclass{aastex63}

\def\cof {$^{12}\mathrm{CO}~(J=1\rightarrow0)$}
\def\cos {$^{13}\mathrm{CO}~(J=1\rightarrow0)$}
 \def\cotss{$^{12}\mathrm{CO}~(J=3\rightarrow2)$}

\def\cot {$\mathrm{C}^{18}\mathrm{O}~(J=1\rightarrow0)$}

\def\cofs {$^{12}\mathrm{CO}$}
\def\coss {$^{13}\mathrm{CO}$}
\def\cots{$\mathrm{C}^{18}\mathrm{O}$}
\def\deg  {\ifmmode {^\circ}\else {$^\circ$}\fi}

\def\kms     {km~s$^{-1}$}
\newcommand{\HI}{\mbox{H\,\textsc{i}}}%
\newcommand{\HII}{\mbox{H\,\textsc{ii}}}%

\usepackage{booktabs}
\usepackage{chngcntr}

\date{\today}
\submitjournal{ApJ}

\shorttitle{Beam filling factors}
\shortauthors{Yan et al.} 

\begin{document}

\title{On the Beam Filling Factors of Molecular Clouds}

\correspondingauthor{Ji Yang}
\email{jiyang@pmo.ac.cn,qzyan@pmo.ac.cn}

\author[0000-0003-4586-7751]{Qing-Zeng Yan}
\affil{Purple Mountain Observatory and Key Laboratory of Radio Astronomy,\\
 Chinese Academy of Sciences, 10 Yuanhua Road, Qixia District, Nanjing 210033, People's Republic of China}

\author[0000-0001-7768-7320]{Ji Yang}
\affil{Purple Mountain Observatory and Key Laboratory of Radio Astronomy,\\
 Chinese Academy of Sciences, 10 Yuanhua Road, Qixia District, Nanjing 210033, People's Republic of China}

 \author[0000-0002-0197-470X]{Yang Su }
 \affil{Purple Mountain Observatory and Key Laboratory of Radio Astronomy,\\
 Chinese Academy of Sciences, 10 Yuanhua Road, Qixia District, Nanjing 210033, People's Republic of China}

 \author[0000-0002-3904-1622]{Yan Sun}
\affil{Purple Mountain Observatory and Key Laboratory of Radio Astronomy,\\
 Chinese Academy of Sciences, 10 Yuanhua Road, Qixia District, Nanjing 210033, People's Republic of China}
 
  \author[0000-0001-8923-7757]{Chen Wang}
 \affil{Purple Mountain Observatory and Key Laboratory of Radio Astronomy,\\
 Chinese Academy of Sciences, 10 Yuanhua Road, Qixia District, Nanjing 210033, People's Republic of China}





\begin{abstract}
Imaging surveys of CO and other molecular transition lines are fundamental to measuring the large-scale distribution of molecular gas in the Milky Way. Due to finite angular resolution and sensitivity, however, observational effects are inevitable in the surveys, but few studies are available on the extent of uncertainties involved. The purpose of this work is to investigate the dependence of observations on angular resolution (beam sizes), sensitivity (noise levels), distances, and molecular tracers. To this end, we use high-quality CO images of a large-scale region ($25\fdg8  <l<49\fdg7 $ and $|b|<5\deg$) mapped by the Milky Way Imaging Scroll Painting (MWISP) survey as a benchmark to simulate observations with larger beam sizes and higher noise levels, deriving corresponding beam filling and sensitivity clip  factors. The sensitivity clip factor is defined to be the completeness of observed flux.  Taking the entire image as a whole object, we found that \cofs\ has the largest beam filling and sensitivity clip factors and \cots\ has the lowest. For molecular cloud samples extracted from images, the beam filling factor can be described by a characteristic size, $l_{1/4}=0.762$ (in beam size), at which the beam filling factor is approximately 1/4. The sensitivity clip factor shows a similar relationship but is more correlated with the mean voxel signal-to-noise ratio of molecular clouds.  This result may serve as a practical reference on beam filling and sensitivity clip factors in further analyses of the MWISP data and other observations. 
\end{abstract}



\section{Introduction} \label{sec:intro}

Molecular clouds are a kind of neutral interstellar medium (ISM) \citep{2015ARA&A..53..583H}, characterized with low temperatures \citep{1983A&A...128..212M} and relatively high densities \citep{2001ApJ...547..792D}. In terms of morphology, molecular clouds are clumpy \citep{1980ApJ...238..158N} with fractal boundaries \citep{1991ApJ...378..186F, 1998A&A...336..697S, 1999MNRAS.302..417S}, and many show filamentary structures at large scale \citep{1987ApJ...312L..45B, 2010A&A...518L.100M, 2010A&A...518L.102A}. Having this particular kind of structure, the surface brightness temperature of molecular clouds is inhomogeneous \citep{1990ApJ...365..620B},  causing non-unity beam filling factors subjected to observations with finite beam sizes and sensitivities. However, beam filling factors are usually assumed to be unity in the calculation of physical properties, such as the excitation temperature, the optical depth, and the mass,  which is inaccurate and may introduce systematic errors.

 Observationally, the beam filling factor appears to be an item of diminishing the brightness temperature \citep[e.g.,][]{2015PASP..127..266M}. In the low temperature approximation, the specific form of the radiative equation is 
\begin{equation}
T_{\rm mb} = \eta \left(T_{\rm ex}-T_{\rm bg}\right)\left(1-\exp\left(-\tau\right)\right),
\label{eq:bffradiation}
\end{equation}
where $T_{\rm mb}$ is the observed brightness temperature,  $\eta$ is the beam filling factor, $T_{\rm ex}$ is the excitation temperature, $T_{\rm bg}$ is the background temperature, and $\tau$ is the optical depth. However, in practical observations, $T_{\rm mb}$ is clipped due to the limited sensitivity, i.e., the brightness temperature below sensitivity is clipped to be zero in the observational data.  We refer to this sensitivity effect (flux completeness) as the sensitivity clip factor, $\xi$.  Obviously, the observed brightness temperature of molecular clouds is an interplay of sensitivity and angular resolution. 

   $\xi$, the sensitivity clip factor, is defined as the ratio of the observed flux and the total flux. The total flux corresponds to the flux at perfect sensitivity, and can be estimated by extrapolation from the observed flux at different sensitivity levels. The observed flux, however, corresponds to the flux at finite sensitivity. 

 Geometrically, the beam filling factor can also be defined by 
\begin{equation}
\eta \approx  \frac{ \Omega_{\rm s}}{ \Omega_{\rm A} },
\label{eq:bffdefinition} 
\end{equation}
where $\Omega_{\rm s}$ is solid angle of objects within the antenna beam and $\Omega_{\rm A}$ is the beam solid angle, respectively. Apparently, for an object with a uniform brightness temperature across $\Omega_{\rm A}$, $\Omega_{\rm s}$  equals $\Omega_{\rm A}$, i.e., $\eta =1$, otherwise $\eta <1$.

Beam filling factor effects are particularly severe in extragalactic observations. For example, \citet{2006PASP..118..590R} studied the bias of Giant molecular cloud (GMC) properties caused by limited resolutions (spatial and spectral) and sensitivities toward galaxies in the Local Group. They found the measurement bias could be more than 40\% and recommended a quadratic extrapolation to correct the flux of molecular clouds. \citet{2018ApJ...860..172S} studied molecular cloud properties in 15 nearby galaxies with spatial resolution of 45-120 pc, and given the size of molecular cloud complexes \citep[$\sim$100 pc;][]{2018ARA&A..56...41M}, the beam filling factors of extragalactic GMCs under Atacama Large Millimeter Array (ALMA) observations may be significantly less than unity. They concluded that the beam filling factor may cause the  virial parameter ($\alpha_{\rm vir}$) to be overestimated, due to the underestimation of molecular cloud mass. \citet{2019A&A...625A.148D} derived a surface beam filling factor of $\sim$0.05 for \cofs\ cloud clumps toward M31 (with 11-pc resolution), and in the  central region, the beam filling factor of dense gas is less than 0.02 with a resolution of about 100 pc \citep{2016A&A...585A..44M}. For some studies,  such as metallicity gradients \citep{2020MNRAS.495.3819A} and radiative transfer analyses subjected to sub-beam structures \citep{2017ApJ...835..217L}, the beam size effect is pivotal.

Observations of molecular clouds in the Milky Way are also not free of beam dilution and sensitivity clip effects, particularly for  molecular clouds with small angular sizes. Due to the inhomogeneity, beam smoothing  diminishes the peak and edge brightness temperatures, and when observed with finite sensitivities and spatial resolutions, both the flux and the brightness temperature are underestimated. For instance, \citet{2020ApJ...898...80Y} found a completeness of 80\% for the observed flux of local molecular clouds in the first Galactic quadrant. This completeness is expected to be lower for molecular clouds in distant spiral arms. In addition to the flux completeness, \citet{2018ApJ...858...16G} found with numerical simulations that the $X_{\rm CO}$ factor would increase by a factor of 2 if the beam size increases by a factor of 100 (from 1 to 100 pc). In addition to molecular clouds, the beam filling factor of \HI\ gas is also less than unity. For instance, \citet{2003ApJ...586.1067H}  derived a value of $\sim$0.5 for the warm  \HI\ gas. The beam filling factor can be the largest error source for analyses of small objects, for example, in the study of molecular outflows \citep{2010MNRAS.409...29F}.

 Conventionally, $\eta$ can be estimated in two ways. The first method uses Equation \ref{eq:bffradiation} and an optically thick spectral line with an assumed excitation temperature, but this is inaccurate due to unknown optical depths and guessed excitation temperatures. The other approach is based on the assumption of Gaussian source distribution \citep{2008A&A...482..197P}. Under this assumption, the beam filling factor is estimated to  be $\Theta_{\rm s}^2/\left(\Theta_{\rm s}^2+\Theta_{\rm b}^2\right)$, where  $\Theta_{\rm b}$ and $\Theta_{\rm s}$ are the beam size and  full width at half maximum (FWHM) of the source, respectively. This is a good approximation for stars and dense cores, but for molecular clouds, whose surface brightness temperature distributions are usually none-Gaussian, the beam filling factor may not follow this convolution approach. 



In this paper, we use images of three CO isotopologue lines in the first Galactic quadrant ($25\fdg8  <l<49\fdg7$, $|b|<5\deg$, and $-79<V_{\rm LSR}$ <139 \kms) to study observational effects on spectroscopic survey data of molecular clouds, including the beam filling and sensitivity clip factors. This region has been mapped by the Milky Way Imaging Scroll Painting (MWISP) CO survey \citep{2019ApJS..240....9S} with high sensitivity ($\sim$0.5 K for \cofs) and medium angular resolution (about 50\arcsec). The high dynamical range of the MWISP survey in scale makes this region a superb data set for studying the beam filling factor. The entire data set is roughly divided into four spiral arms based on their radial velocities, and examinations of beam filling factors are subsequently performed on those arm segments.






This paper is organized as follows. The next section (Section \ref{sec:data}) describes the CO data, cloud identification methods, and the beam filling and sensitivity clip factor models. Section \ref{sec:result} presents results of beam filling  and sensitivity clip factors, including collective and individual molecular clouds. Discussions   are presented in Section \ref{sec:discuss}, and we summarize the conclusions in Section \ref{sec:summary}. 









\section{Data And Methods} 
\label{sec:data}

\begin{deluxetable}{cccccccc}
\tablecaption{Observation parameters of three CO isotopologue lines of the MWISP survey.\label{Tab:lineParameters}}
\tablehead{
\colhead{Tracer} & \colhead{Rest frequency} & \colhead{Effective critical density\tablenotemark{a}} &\colhead{HPBW} & \colhead{$T$$_{\rm sys}$}  & 
\colhead{$\delta v$} & \colhead{rms noise} \\
\colhead{   } & \colhead{(GHz)} & \colhead{($10^3$ cm$^{-3}$)} &\colhead{($''$)} & \colhead{ (K) } &  
\colhead{(\kms)} & \colhead{(K) } 
}

\startdata
\cof  & 115.27 &  $\sim$0.06& 49 &220-300   & 0.158 & $\sim$0.49 \\\
\cos &  110.20   & $\sim$6 & 52    &140-190    & 0.166&  $\sim$0.23 \\
\cot &  109.78   & $\sim$18&   52    &140-190    & 0.167& $\sim$0.23 \\
\enddata
\tablenotetext{a}{The effective critical density takes account of radiative line trapping \citep{2019ApJ...885...19Y}.}
\end{deluxetable}

\subsection{CO data} 
We select a region in the first Galactic quadrant ($25\fdg8  <l<49\fdg7$, $|b|<5\deg$, and $-79 <V_{\rm LSR}< 139$ \kms) to study the beam filling and sensitivity clip  factors, and this region has been uniformly mapped by the MWISP\footnote{\href {http://www.radioast.nsdc.cn/mwisp.php}{http://www.radioast.nsdc.cn/mwisp.php}} CO survey \citep{2019ApJS..240....9S}. Observations were performed with the Purple Mountain Observatory (PMO) 13.7-m millimeter telescope, containing three CO isotopologue line maps, \cof, \cos, and \cot, which were all used in this work.

 The angular resolution of these line maps are approximately 49\arcsec, 52\arcsec, and 52\arcsec, and the velocity resolutions are 0.158, 0.166, and 0.167 \kms, respectively. The pixel size of the regridded map is  30\arcsec.  The  rms noise  of \cofs\ is about 0.49 K, $\sim$0.23 K for \coss\ and \cots.  See Table \ref{Tab:lineParameters} for a summary of the observation parameters.

In order to investigate the beam filling and sensitivity clip effects at different distance layers, we roughly split each of the three isotopologue data cubes into four arm  segments \citep{2016ApJ...823...77R} along the  $V_{\rm LSR}$ axis: (1) the Local arm (-6 to 30 \kms); (2) the Sagittarius arm (30 to 70 \kms); (3) the Scutum arm (70 to 139 \kms); (4) the Outer and Outer Scutum-Centaurus arm ($-79$ to $-6$ \kms).  Based on kinematic distances \citep[A5 model in ][]{2014ApJ...783..130R}, distances of the four spiral arms are approximately 1, 3, 6, and 15 kpc, respectively.  The Perseus arm is largely overlapped with the Local arm in $V_{\rm LSR}$ space, so we ignored the Perseus arm. Consequently, three CO lines collectively yield 12 data cubes.

\subsection{Molecular cloud samples}
 \label{sec:cloudIdentification}

In order to investigate the beam filling and sensitivity clip factors of different molecular cloud species, we use the DBSCAN\footnote{\href {https://scikit-learn.org/stable/modules/generated/sklearn.cluster.DBSCAN.html}{https://scikit-learn.org/stable/modules/generated/sklearn.cluster.DBSCAN.html}} algorithm to draw samples from the position-position-velocity (PPV) cubes \citep{2020ApJ...898...80Y}. DBSCAN ignores internal structures of molecular clouds and identifies independent structures in PPV space, sufficing for the beam filling factor studies.

In PPV space, DBSCAN has two parameters, MinPts and the connectivity. The connectivity (three types in PPV space) defines the neighborhood of each voxel, i.e., whether two voxels are connected. For a given voxel, if the number of its neighboring voxels (including itself) is $\geq$ MinPts, it is a core point, and connecting core points and their neighbors define a molecular cloud.  As discussed in \citet{2020ApJ...898...80Y}, for small MinPts values, the three connectivity types provide similar cloud samples, so we simply use connectivity 1 and MinPts 4. The minimum cutoff of the data cube is 2$\sigma$ ($\sim$1 K for \cofs\ and $\sim$0.5 K for \coss\ and \cots), and in practice, the rms  noise  calculation is accurate to each spectrum. 

We applied the post selection criteria to remove small DBSCAN clusters that are likely to be noise \citep{2020ApJ...898...80Y}. The post selection criteria contain four conditions: (1) the  voxel number is $\geq$ 16; (2) the peak brightness temperature is $\geq$ 5$\sigma$; (3) the projection area contains a beam (a compact 2$\times$2 region equivalent to 60\arcsec$\times$60\arcsec); (4) the velocity channel number is $\geq$ 3. 


 \subsection{Beam filling and sensitivity clip factors}

 The beam filling   and  sensitivity clip factors have different applications. The beam filling factor is used to correct $T_{\rm mb}$ to obtain accurate excitation temperatures and optical depths, while  the sensitivity clip factor is used to correct the observed flux, which is more related to, e.g., the mass of molecular clouds. $\eta$ strongly depends on the beam size, and we refer to $\eta$ as the value at the angular resolution of the data. For a single pixel, $\xi$ is either unity or zero, but for an image or a molecular cloud, the observed flux above cutoffs is the inverse cumulative distribution function of CO brightness temperatures, and $\xi$ describes the observed fraction of the flux.

 $\eta$ and $\xi$ can be estimated in two approaches: (1) based on the entire image and (2) based on molecular cloud samples. In the first image-based case, the whole data cube is taken as a single molecular cloud, while in the second sample-based case, the estimation is performed for each molecular cloud sample identified with the method described in Section \ref{sec:cloudIdentification}.


 The variation of $\eta$ and $\xi$ is modeled with extrapolation functions.  Without a physically motivated theory at hand, we use an empirical function. However, the extrapolation function should be simple and versatile, applicable to both image-based and sample-based cases.  
 
 In order to model $\eta$ and $\xi$, we produce two data sets based on the MWISP CO data. The first data set simulates a series of observations at different beam sizes and is used to estimate $\eta$. For the convenience of calculation, we keep the pixel size constant in smoothing. The second data set, however, resembles observations with the same beam size but different sensitivity clips and is used to estimate $\xi$.



\subsubsection{Beam filling factors}

For an isolated Gaussian source, the variation of its peak brightness temperature is proportional to the beam filling factor, but for molecular clouds, which are irregular and  have non-uniform brightness distribution, we can take each voxel in a data cube as a Gaussian peak. In this case, the beam filling factor can be examined voxel by voxel based on intensity, but the signal-to-noise ratio (SNR) of a single voxel is low, causing large errors in curve fitting, particularly for extended weak components of molecular clouds. In order to obtain high SNRs, we take each molecular cloud as an object and use the mean $T_{\rm mb}$ to derive an average beam filling factor.


For specific observation data, the rms noise and the angular resolution are coupled. Smoothing operations decrease both $T_{\rm mb}$ and  the rms noise, and in the estimation of  $\eta$, voxels are need to be above the sensitivity level in all simulated observations. In other words, voxels that are below the sensitivity level in a smoothing case are discarded. The sensitivity level we used is 2$\sigma$, consistent with DBSCAN parameters. However, $\sigma$ is different between smoothing cases.

The procedure of obtaining $\eta$ contains three main steps: (1) identifying molecular clouds, (2) smoothing data cubes, and (3) modeling $\eta$. 
The first step applies the procedure of producing molecular cloud samples (see Section \ref{sec:cloudIdentification}) on raw CO data.  

In the second step, data cubes are smoothed to simulate observations with larger beam sizes. The smoothing operation is performed with the \texttt{spectral-cube}  package\footnote{\href {https://spectral-cube.readthedocs.io/en/latest/index.html}{https://spectral-cube.readthedocs.io/en/latest/index.html}} in  \texttt{Python} language. The beam size varies by factors from 1.5 to 10 with an interval of 0.5, giving 18 smoothing cases in total. For the convenience of comparing, we keep the voxel size unchanged.  The rms noise is calculated with the Outer arm cubes, which contain the largest amount of noise voxels, and we use the rms of negative values in the spectra as a proxy of the rms noise. 
 
In the third step, we estimate $\eta$ based on the variation of mean $T_{\rm mb}$ with respect to the beam size. The mean $T_{\rm mb}$ is obtained by averaging   brightness temperatures over voxels that are above the sensitivity clip levels in all smoothing cases. $\eta$ is obtained through extrapolation, and taking the mean $T_{\rm mb}$ at the zero-beam point as observations with infinite angular resolutions ($\eta=1$), the fraction of $T_{\rm mb}$ at a specific beam size is the corresponding beam filling factor. For MWISP molecular clouds, we use the fraction at the MWISP beam size as their beam filling factors.

We found that the mean $T_{\rm mb}$ roughly contains two components, a linear part and an exponential part, which can be well described by a four-parameter function:
\begin{equation}
T\left(\Theta\right)=a\exp\left(-b\Theta\right)-c\Theta+d,
\label{eq:fffunction}
\end{equation}
where, $\Theta$ represents the beam size, y is the corresponding observed flux, and $a$, $b$, $c$, and $d$ are four parameters to be determined. Equation \ref{eq:fffunction} is approximately linear when $b$ is small and is also able to fit flux variations that decrease fast (with large $b$ values). The superiority of this function over polynomials is that the meaning of Equation \ref{eq:fffunction} is more clear,  and Equation \ref{eq:fffunction} is a monotonic function, which satisfies the intuition that flux decreases with larger beam sizes.





We use Equation \ref{eq:fffunction} to extrapolate the value of the mean $T_{\rm mb}$ to zero. The variation of the mean $T_{\rm mb}$ is fitted with \texttt{curve\_fit} in the \texttt{Python} package \texttt{SciPy} with flux errors considered. The error of the mean $T_{\rm mb}$ is estimated with $\sqrt {\sum_i\sigma_i^2}/N$ , where $N$ is the voxel number and $\sigma_i$ is the rms noise of each voxel. $\sigma_i$ decreases with beam sizes but $N$ is constant. Specifically, the beam filling factor is estimated with Equation \ref{eq:fffunction} using
\begin{equation}
\eta =\frac{T\left(\Theta_{\mathrm{MWISP}}\right)}{T\left(0\right)},
\label{eq:ffmwisp}
\end{equation}
where  $T\left(\Theta_{\mathrm{MWISP}}\right)$ and $T\left(0\right)$ is the mean $T_{\rm mb}$ at the MWISP beam size and at zero-beam size, respectively. Errors of $T\left(\Theta_{\mathrm{MWISP}}\right)$  and $T\left(0\right)$ are obtained with first derivatives of $T$ at $\Theta_{\mathrm{MWISP}}$ and 0, respectively, together with the covariance of $a$, $b$, and $c$, and the error of $\eta$ is subsequently estimated with propagation of errors.

\subsubsection{Sensitivity clip factors}

 In this section, we present the method of deriving $\xi$. We use the cutoff as a proxy of the sensitivity clip levels, simulating observations at different sensitivities but with the same angular resolution. In this context, the flux above the cutoff is the inverse of the cumulative distribution function of the brightness temperature. 



The procedure of modeling $\xi$ is similar to that of $\eta$. By definition, $\xi$ approaches unity as the sensitivity goes infinity, and $\xi$ corresponds to the completeness of the flux at a specific sensitivity level. The cutoffs range from 2$\sigma$ to 20$\sigma$ with an interval of 0.2$\sigma$. $\xi$ is the fraction of observed flux at 2$\sigma$ with respect to the zeroth flux obtained with extrapolation.

Equation \ref{eq:fffunction} cannot model the flux variation with respect to the sensitivity (the brightness temperature cutoff). Instead, we found that the quadratic equation suggested by \citet{2006PASP..118..590R} is more appropriate. However, toward high cutoffs, the observed flux of molecular clouds usually decreases rapidly to zero, so we use a sigmoid term that contains the Gaussian CDF to model this zero tail.

specifically, the observed flux ($f$) above the cutoff ($x$) is approximately 
\begin{equation}\label{eq:modelsens}
 \left \{\begin{array}{ll}
 \mathrm{erf}(z)= \frac{2}{\sqrt\pi}\int_0^z\exp(-t^2)dt,\\
F\left(x\right) = \frac{1}{2} \left( 1+ \mathrm{erf}\left(\frac{x-\mu}{\sqrt2\delta}\right)  \right),\\
f\left(x\right)= \left(a(x-b)^2+c\right)\left(1-  F\left(x\right)   \right),
\end{array}
\right.
\end{equation}
where $F\left(x\right)$ is the cumulative distribution function (CDF) of Gaussian distribution $\mathcal{N}\left(\mu,\delta\right)$, $x$ is in units of rms noise ($\sigma$), and $f\left(x\right)$ is the observed flux above $x$. Due to the sigmoid item that contains the Gaussian CDF, $f\left(x\right)$ is forced to approximate 0 for large $x$ values. In total, Equation \ref{eq:modelsens} contains 5 parameters: $a$, $b$, $c$, $\mu$, and $\delta$, and for a normal fitting, $a$, $b$, and $\delta$ should be positive.  $\xi$ is estimated subsequently with
\begin{equation}
\xi =\frac{f\left(\sigma_{\rm cut}\right)}{f\left(0\right)},
\label{eq:ffmwispcut} 
\end{equation}
where $\sigma_{\rm cut}$ is the $T_{\rm mb}$ cutoff (in the unit of rms noise) of molecular clouds.

\begin{deluxetable*}{ccccccccccccccccccccc} 
\tabletypesize{\scriptsize}
 \setlength{\tabcolsep}{2.5pt}
\tablecaption{Variation of the mean $T_{\rm mb}$ with respect to beam sizes for molecular clouds in four Galactic arm segments. \label{tab:beamflux}}
\tablehead{
\colhead{ }  & \colhead{ } & \multicolumn{19}{c}{Beam sizes} \\ 
\cline{3-21}
\colhead{Arm}  & \colhead{Line} & \colhead{1}  & \colhead{1.5} & \colhead{2} & \colhead{2.5} & \colhead{3}  & \colhead{3.5}  & \colhead{4} 
& \colhead{4.5}  & \colhead{5} & \colhead{5.5} & \colhead{6} & \colhead{6.5}  & \colhead{7}  & \colhead{7.5} 
& \colhead{8}  & \colhead{8.5} & \colhead{9} & \colhead{9.5} & \colhead{10}  \\
\cline{3-21}
\colhead{ }  & \colhead{ } &  \multicolumn{19}{c}{ ($\rm K$) }
} 
\startdata  
 & \cofs & 2.55 & 2.46 & 2.42 & 2.39 & 2.37 & 2.35 & 2.33 & 2.31 & 2.29 & 2.28 & 2.26 & 2.25 & 2.23 & 2.22 & 2.21 & 2.19 & 2.18 & 2.17 & 2.16  \\
Local & \coss & 1.22 & 1.15 & 1.12 & 1.10 & 1.08 & 1.06 & 1.05 & 1.03 & 1.02 & 1.01 & 0.996 & 0.985 & 0.975 & 0.965 & 0.955 & 0.946 & 0.937 & 0.928 & 0.920  \\
 & \cots & 0.950 & 0.836 & 0.792 & 0.762 & 0.736 & 0.713 & 0.692 & 0.672 & 0.654 & 0.637 & 0.622 & 0.607 & 0.593 & 0.580 & 0.567 & 0.556 & 0.544 & 0.534 & 0.524  \\
\hline
 & \cofs & 2.72 & 2.61 & 2.56 & 2.52 & 2.48 & 2.45 & 2.42 & 2.40 & 2.37 & 2.35 & 2.32 & 2.30 & 2.28 & 2.26 & 2.24 & 2.23 & 2.21 & 2.19 & 2.18  \\
Sagittarius & \coss & 1.22 & 1.12 & 1.08 & 1.04 & 1.01 & 0.986 & 0.961 & 0.938 & 0.917 & 0.897 & 0.878 & 0.861 & 0.844 & 0.829 & 0.814 & 0.800 & 0.787 & 0.774 & 0.762  \\
 & \cots & 0.867 & 0.724 & 0.657 & 0.608 & 0.566 & 0.529 & 0.497 & 0.468 & 0.442 & 0.419 & 0.398 & 0.378 & 0.361 & 0.344 & 0.329 & 0.316 & 0.303 & 0.291 & 0.280  \\
\hline
 & \cofs & 3.06 & 2.96 & 2.92 & 2.88 & 2.85 & 2.82 & 2.80 & 2.77 & 2.75 & 2.73 & 2.71 & 2.69 & 2.67 & 2.65 & 2.63 & 2.62 & 2.60 & 2.59 & 2.57  \\
Scutum & \coss & 1.29 & 1.20 & 1.16 & 1.12 & 1.09 & 1.06 & 1.04 & 1.02 & 0.994 & 0.974 & 0.955 & 0.938 & 0.921 & 0.906 & 0.892 & 0.878 & 0.865 & 0.852 & 0.840  \\
 & \cots & 0.914 & 0.765 & 0.694 & 0.642 & 0.598 & 0.560 & 0.527 & 0.497 & 0.471 & 0.448 & 0.426 & 0.407 & 0.390 & 0.374 & 0.359 & 0.346 & 0.333 & 0.322 & 0.311  \\
\hline
 & \cofs & 1.93 & 1.68 & 1.53 & 1.41 & 1.30 & 1.21 & 1.13 & 1.06 & 0.998 & 0.942 & 0.891 & 0.844 & 0.802 & 0.764 & 0.729 & 0.696 & 0.666 & 0.638 & 0.613  \\
Outer & \coss & 0.924 & 0.753 & 0.660 & 0.590 & 0.533 & 0.486 & 0.446 & 0.412 & 0.382 & 0.356 & 0.332 & 0.312 & 0.293 & 0.276 & 0.261 & 0.247 & 0.234 & 0.223 & 0.212  \\
 & \cots & 0.000 & 0.000 & 0.000 & 0.000 & 0.000 & 0.000 & 0.000 & 0.000 & 0.000 & 0.000 & 0.000 & 0.000 & 0.000 & 0.000 & 0.000 & 0.000 & 0.000 & 0.000 & 0.000  \\
\enddata 
\tablecomments{The beam size is in units of the MWISP beam, 49\arcsec\ for \cofs\ and 52\arcsec\ for \coss\ and \cots. }
\end{deluxetable*} 


\begin{deluxetable*}{ccccc} 
\tabletypesize{\scriptsize}

\tablecaption{Image-based beam filling and sensitivity clip factors of the MWISP survey. \label{tab:mwispfilling}}
\tablehead{
\colhead{Arm} & \colhead{Line}& \colhead{$V_{\mathrm{LSR}}$} &       \colhead{ $\eta$ } &     \colhead{ $\xi$ }  
} 
\colnumbers
\startdata 
 & \cofs &  & 0.982 $\pm$ 0.000 & 0.766 $\pm$ 0.000   \\
Local & \coss & [  0, 30] & 0.959 $\pm$ 0.000 & 0.676 $\pm$ 0.000   \\
 & \cots &  & 0.877 $\pm$ 0.001 & 0.450 $\pm$ 0.000   \\
\hline
 & \cofs &  & 0.980 $\pm$ 0.000 & 0.788 $\pm$ 0.000   \\
Sagittarius & \coss & [ 30, 70] & 0.941 $\pm$ 0.000 & 0.644 $\pm$ 0.000   \\
 & \cots &  & 0.722 $\pm$ 0.001 & 0.458 $\pm$ 0.001   \\
\hline
 & \cofs &  & 0.987 $\pm$ 0.000 & 0.834 $\pm$ 0.000   \\
Scutum & \coss & [ 70,139] & 0.956 $\pm$ 0.000 & 0.702 $\pm$ 0.000   \\
 & \cots &  & 0.713 $\pm$ 0.000 & 0.480 $\pm$ 0.000   \\
\hline
 & \cofs &  & 0.777 $\pm$ 0.000 & 0.540 $\pm$ 0.000   \\
Outer & \coss & [-79, -6] & 0.613 $\pm$ 0.001 & 0.412 $\pm$ 0.001   \\
 & \cots & -- & -- & --   \\
\enddata
\end{deluxetable*}

 \begin{figure}[ht!]
 \gridline{\fig{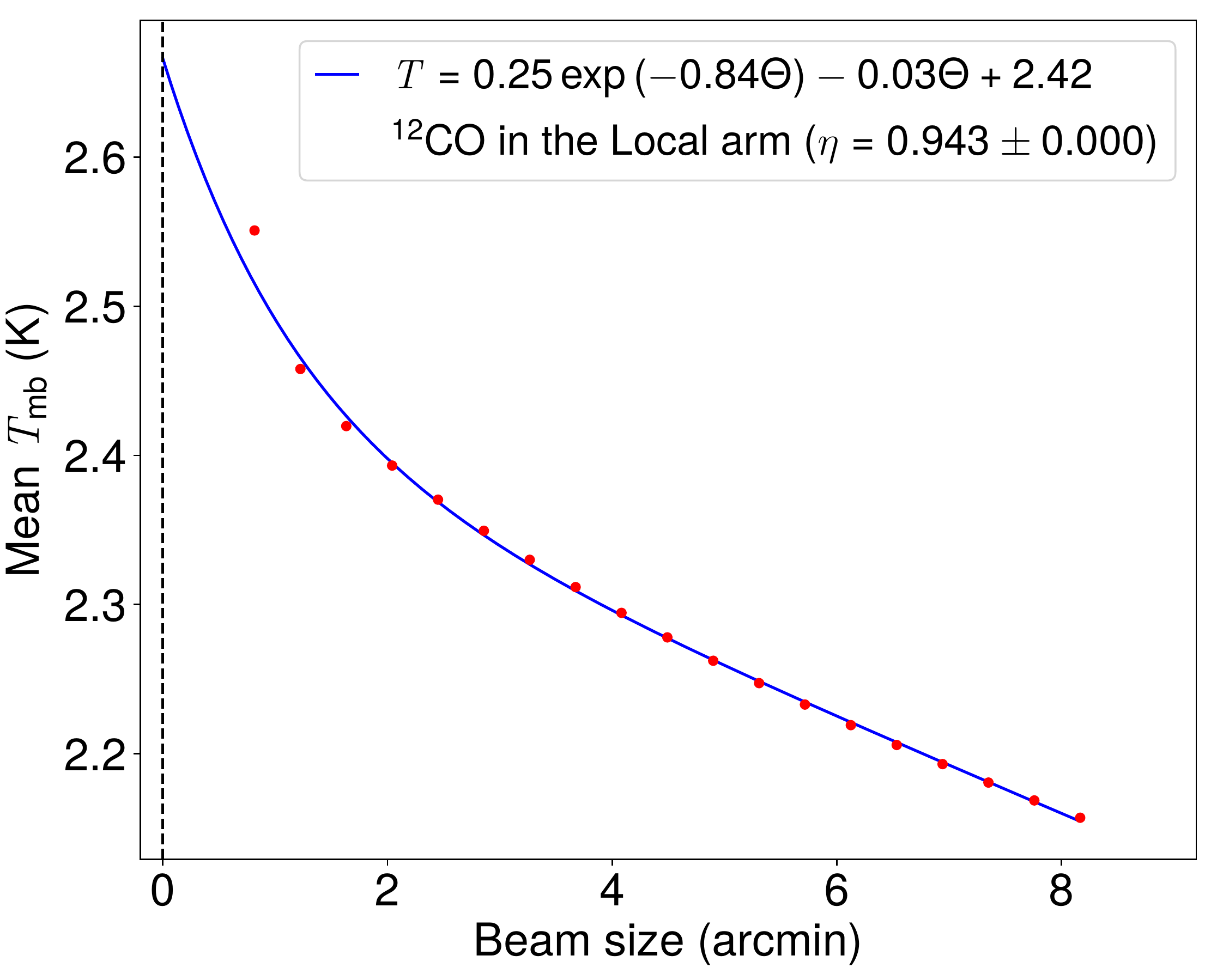}{0.33\textwidth}{(a)  }  \fig{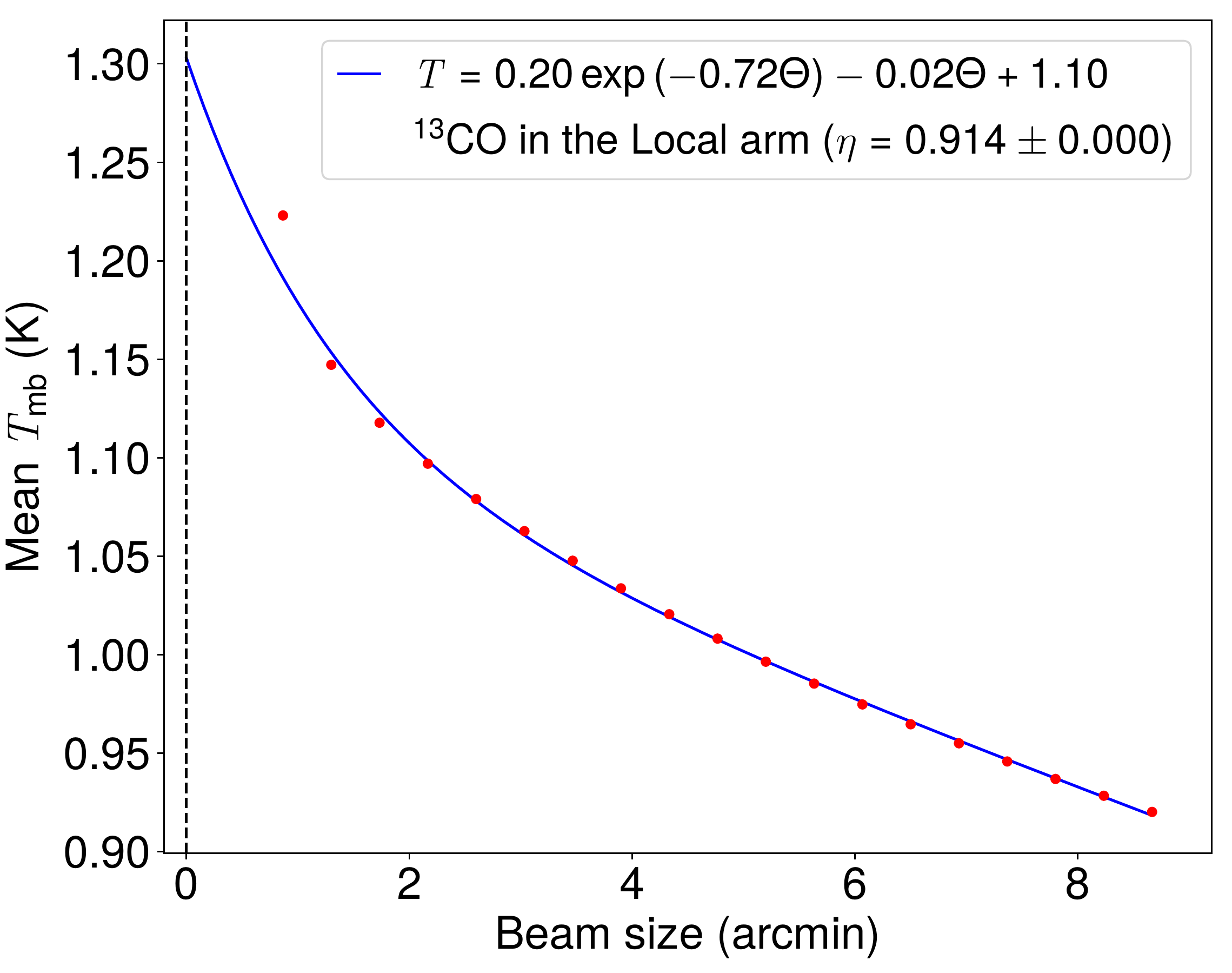}{0.33\textwidth}{(b)  }   \fig{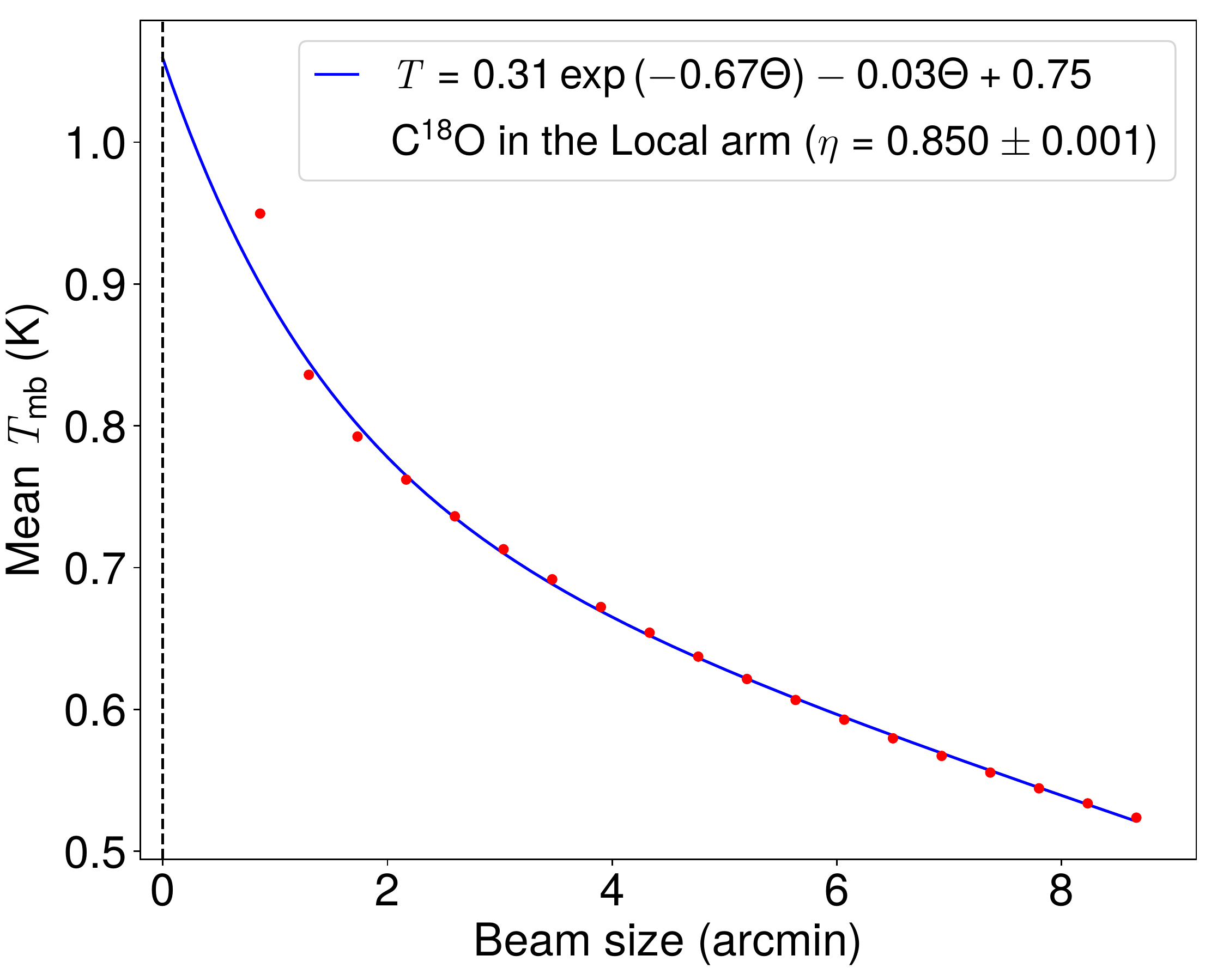}{0.33\textwidth}{(c)    } 
 } 
\caption{  Variation of the mean $T_{\rm mb}$ against the beam size. Only local molecular clouds are displayed:  (a) \cofs, (b) \coss, and (c) \cots. Red points are the observed flux for each smoothing case with different beam sizes, and the error bar is smaller than the marker size. The blue lines are fitted with red points using Equation \ref{eq:fffunction}. \label{fig:fflocal} } 
\end{figure}

\section{Results}
\label{sec:result}

\subsection{ Image-based  beam filling factors} 
 In this section, we demonstrate the results of imaged-based beam filling factors.  Image-based  beam filling factors are calculated by taking the whole data cube as a single molecular cloud. The observed mean $T_{\rm mb}$  toward four spiral arm segments is listed in Table \ref{tab:beamflux}, including all smoothing cases. No \cots\ emission is detected toward the Outer arm, i.e., the beam filling factor of \cots\ in the Outer arm is approximately zero.




As examples, we display  variations of the mean $T_{\rm mb}$ and the image-base  $\eta$ of local molecular clouds in Figure \ref{fig:fflocal}. The relative errors are  small, about $1\times10^{-4}$. The patterns of the mean $T_{\rm mb}$ variations are similar for three CO lines, and Equation \ref{eq:fffunction} fits the variation of the mean $T_{\rm mb}$ well,  except a slight deviation for the mean $T_{\rm mb}$ of the raw data. This systematic shift of the mean $T_{\rm mb}$ between raw and smoothing data is discussed in Section \ref{sec:smdata}.  As expected, \cofs\ has the highest $\eta$, while \cots\ has the lowest. 


Table \ref{tab:mwispfilling} summarizes  $\eta$ of three CO lines in four spiral arm segments. $\eta$ of \cofs\ and \coss\ in the Local, Sagittarius, and Scutum arm are approximately unity, while $\eta$ of \cots\ are significantly lower. In the Outer arm, however, both \cofs\ and \coss\ have low beam filling factors.


 \begin{figure}[ht!]
 \gridline{
 \fig{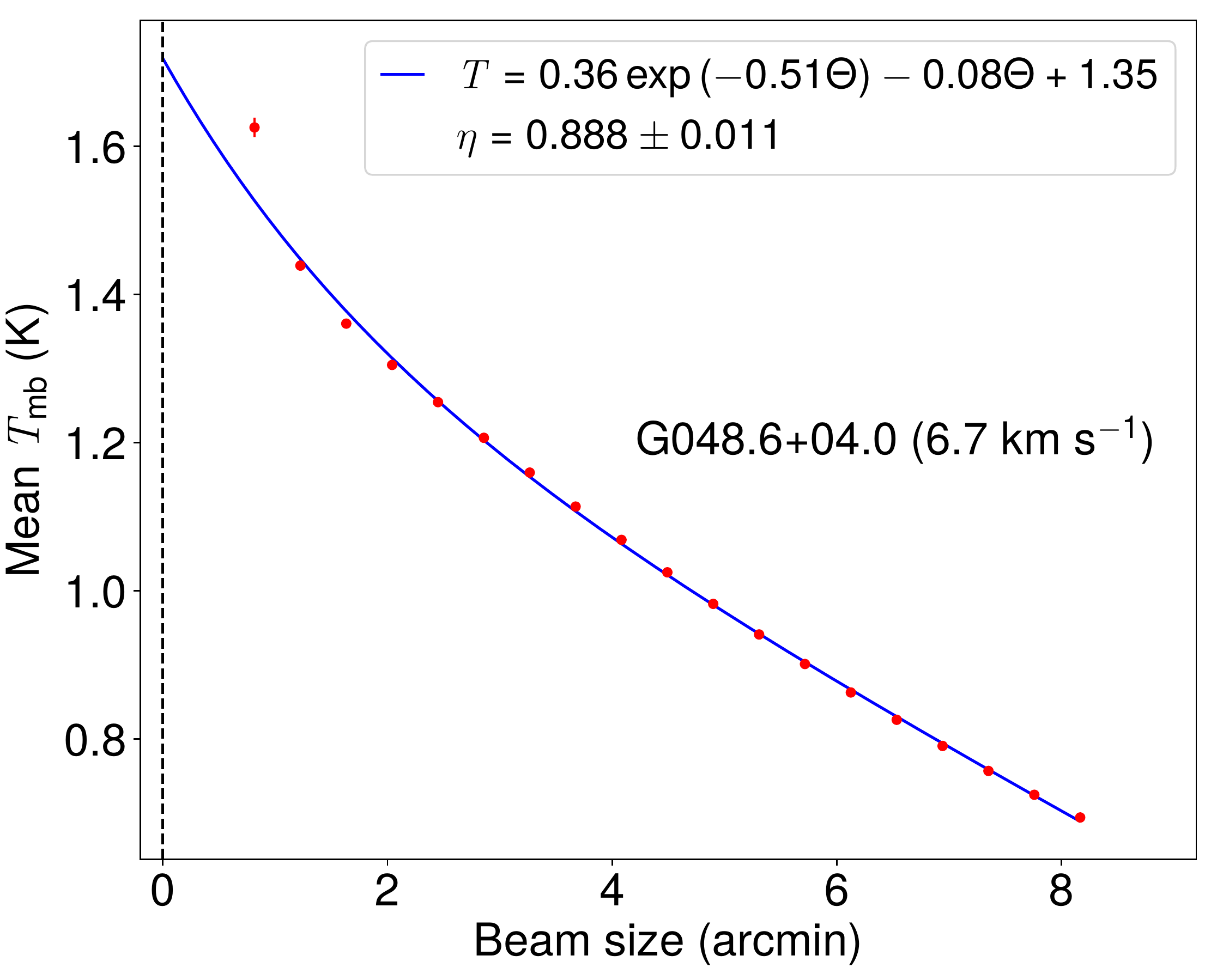}{0.45\textwidth}{(a) } 
 \fig{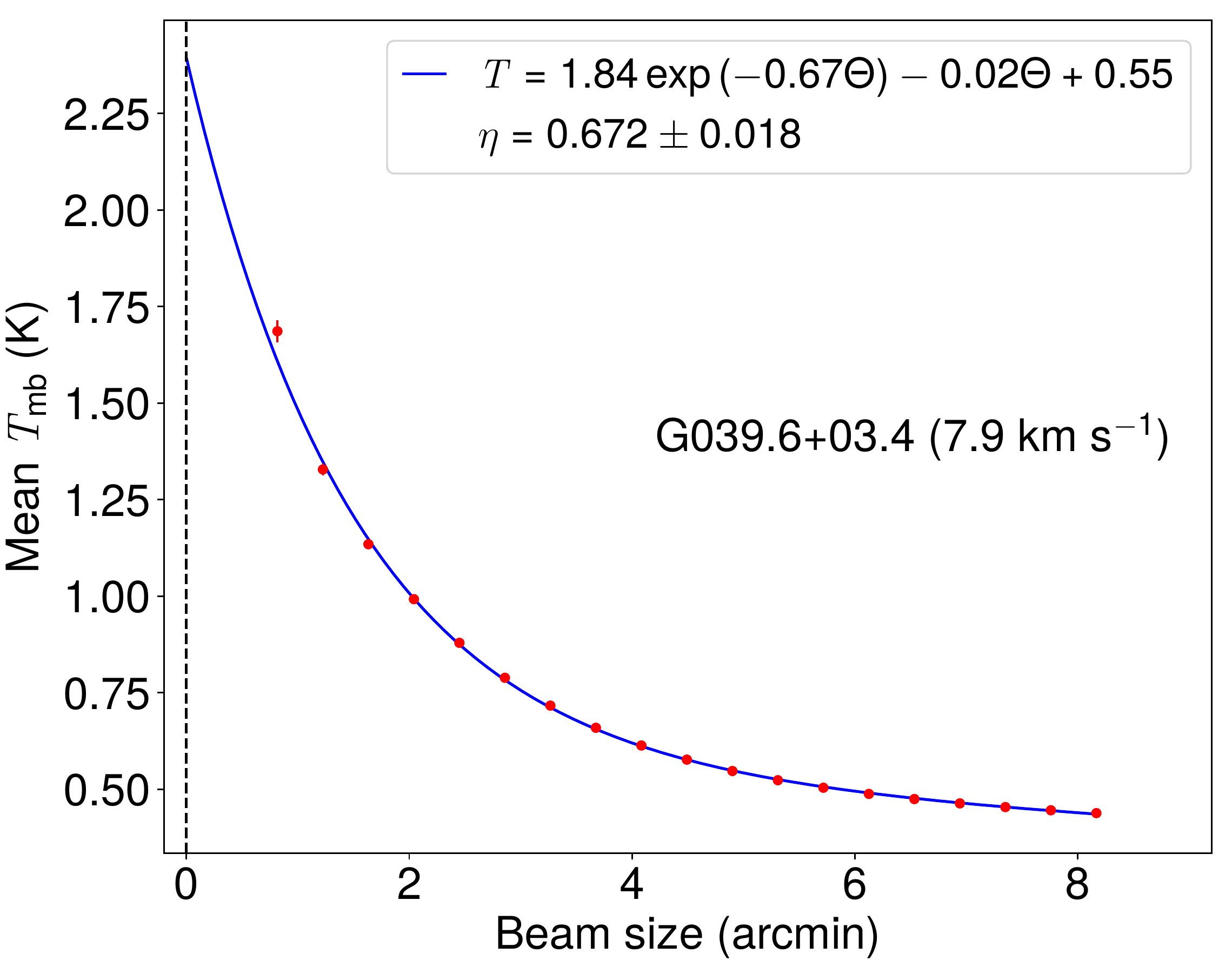}{0.45\textwidth}{(b)}  
 }
 \gridline{
   \fig{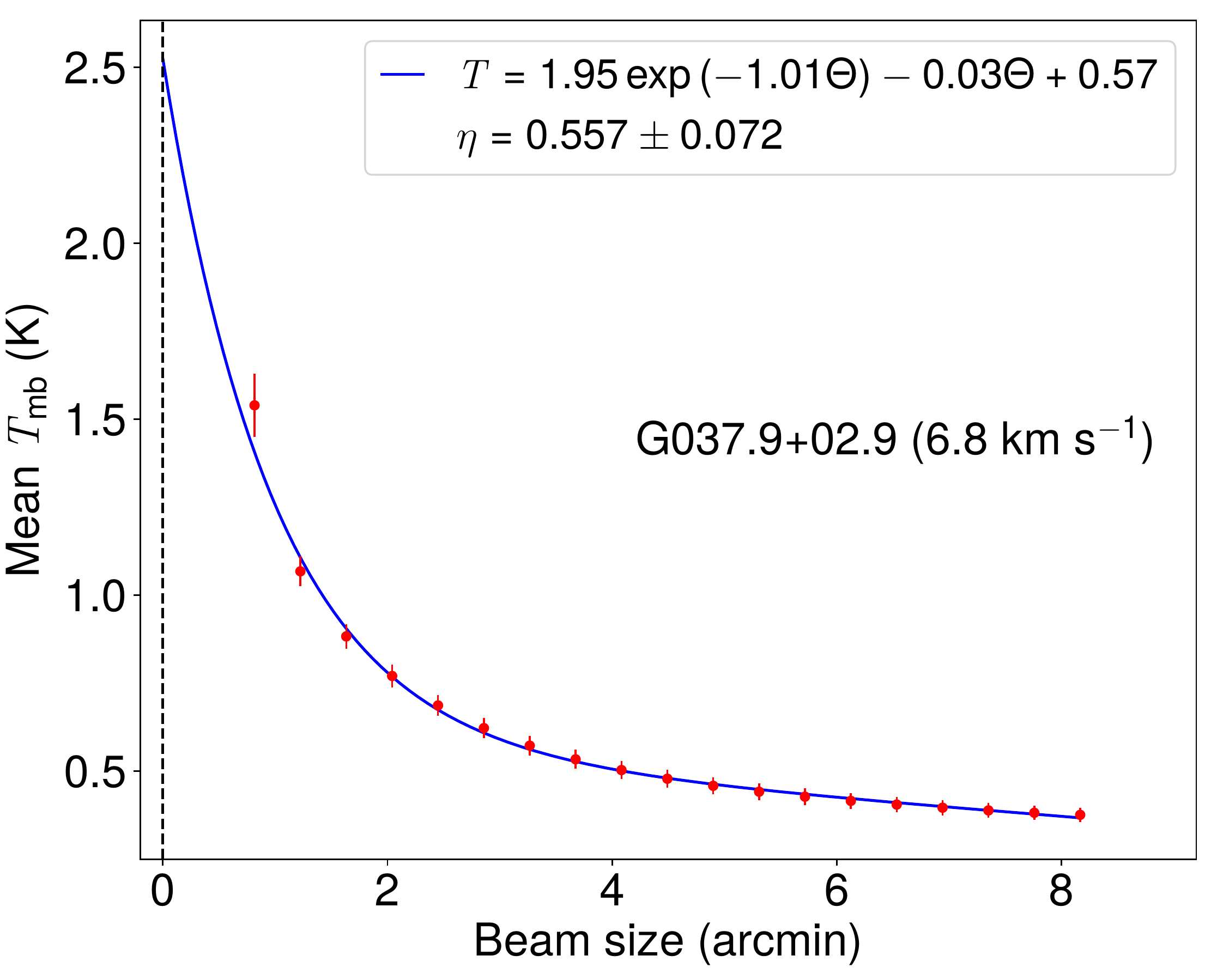}{0.45\textwidth}{(c)}  
   \fig{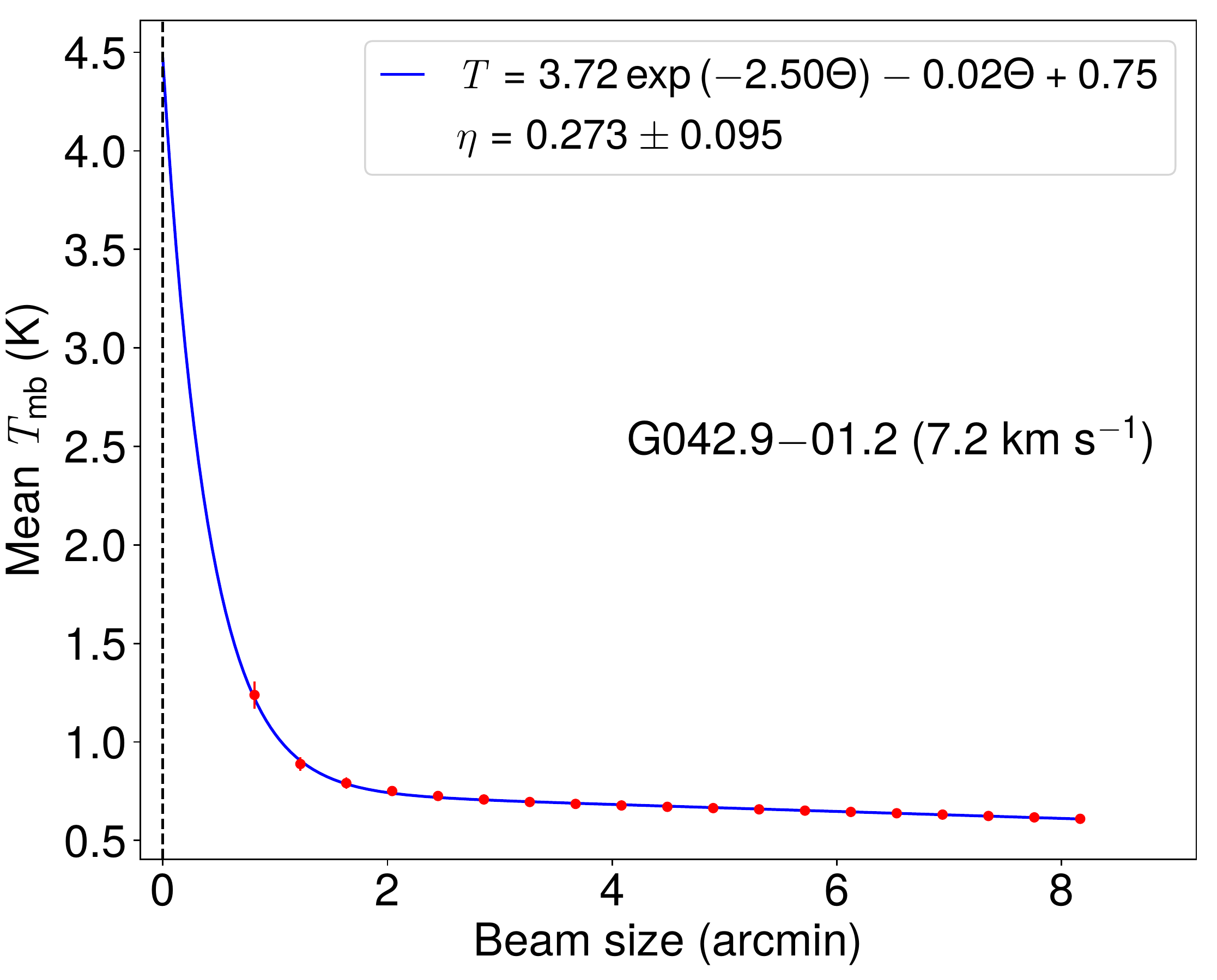}{0.45\textwidth}{(d)} 
 } 
\caption{ Same as Figure \ref{fig:fflocal} but for four typical \cofs\ clouds in the Local arm: (a) G048.6$+$04.0 at 6.7 \kms, (b) G039.6$+$03.4 at 7.9 \kms, (c) G037.9$+$02.9 at 6.8 \kms, and (d) G042.9$-$01.2 at 7.2 \kms.  \label{fig:bfffour} } 
\end{figure}

 \begin{figure}[ht!]
 \gridline{
 \fig{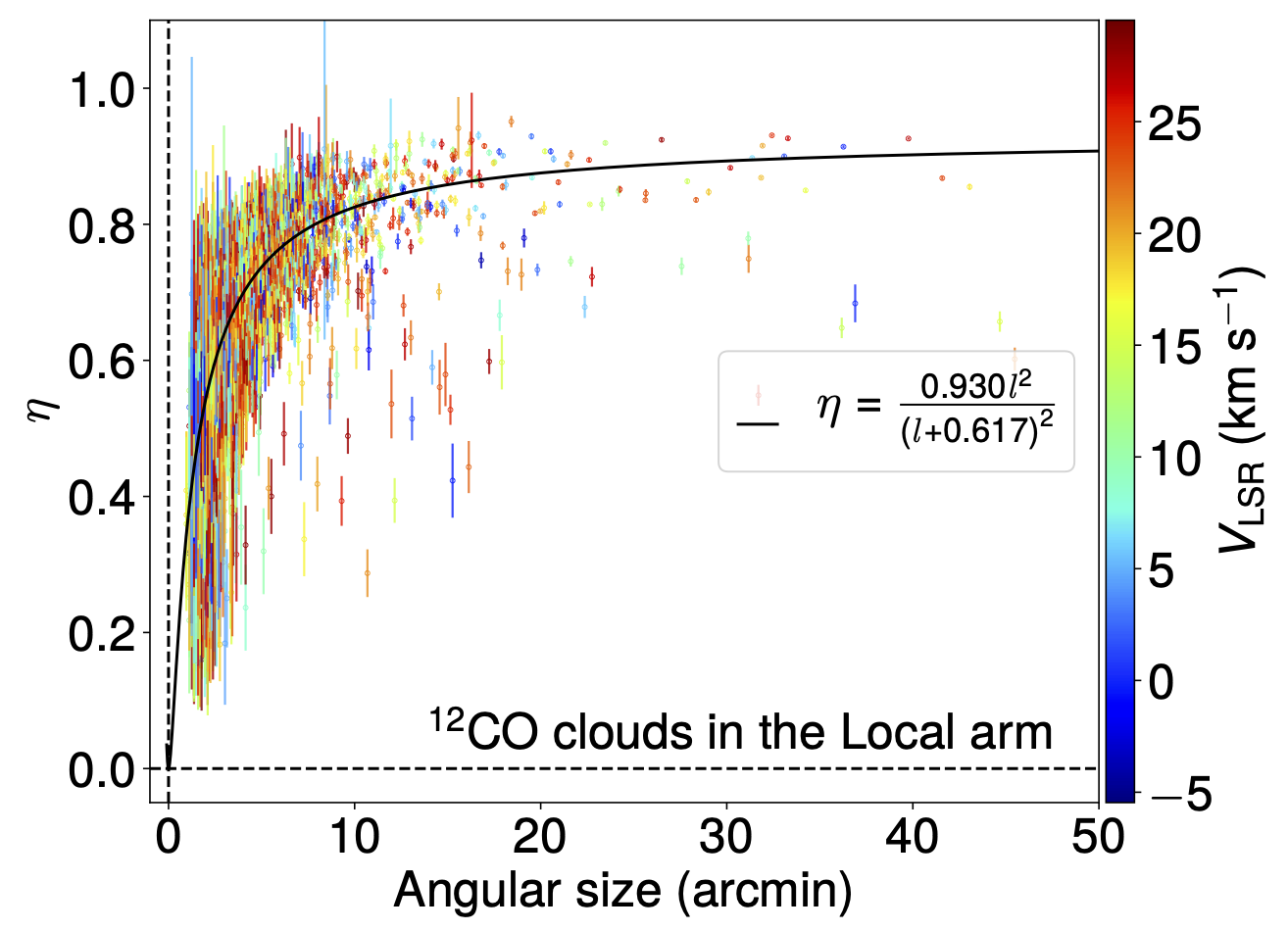}{0.46\textwidth}{(a) } 
 \fig{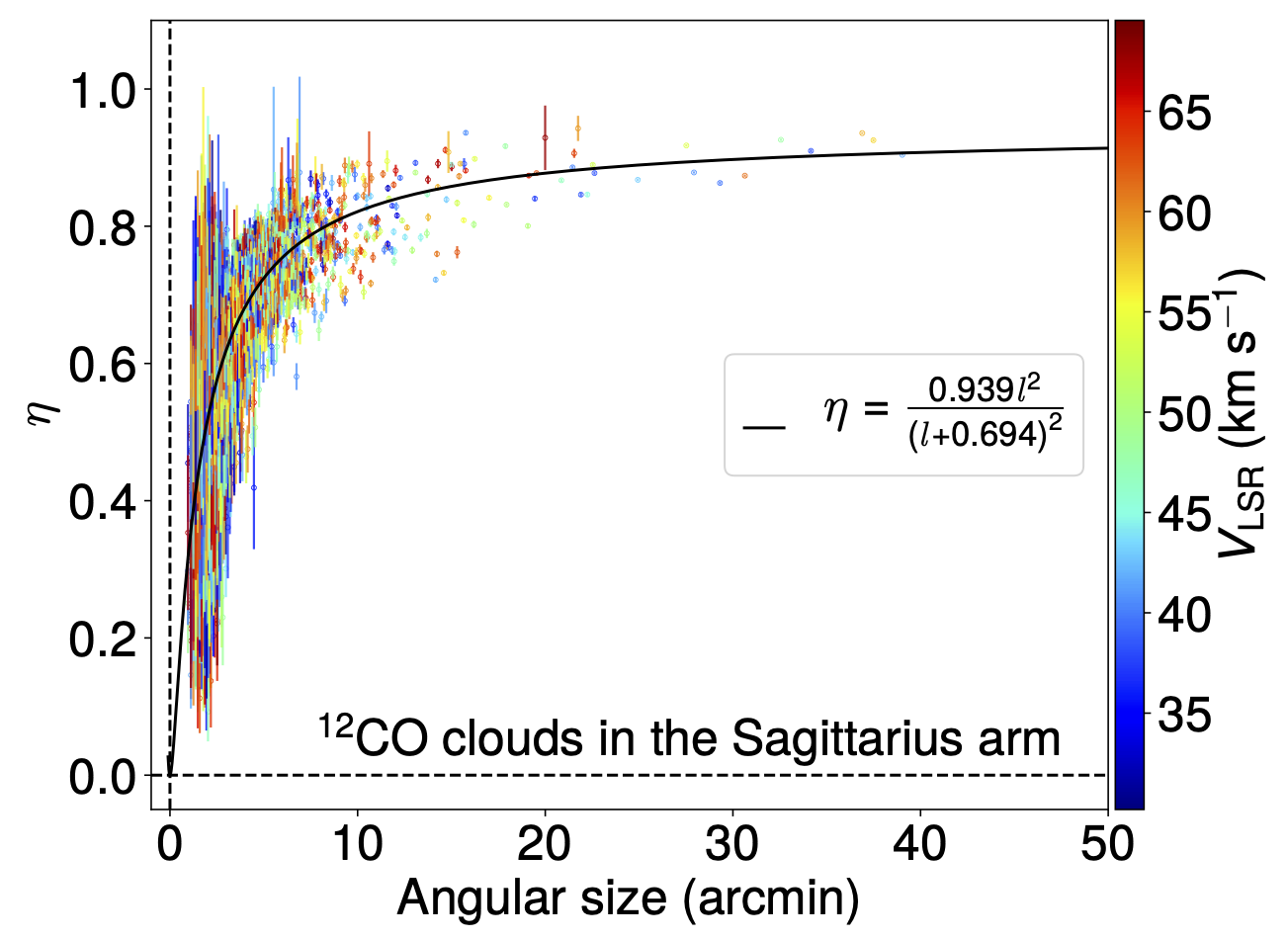}{0.46\textwidth}{(b)}  
 }
 \gridline{
   \fig{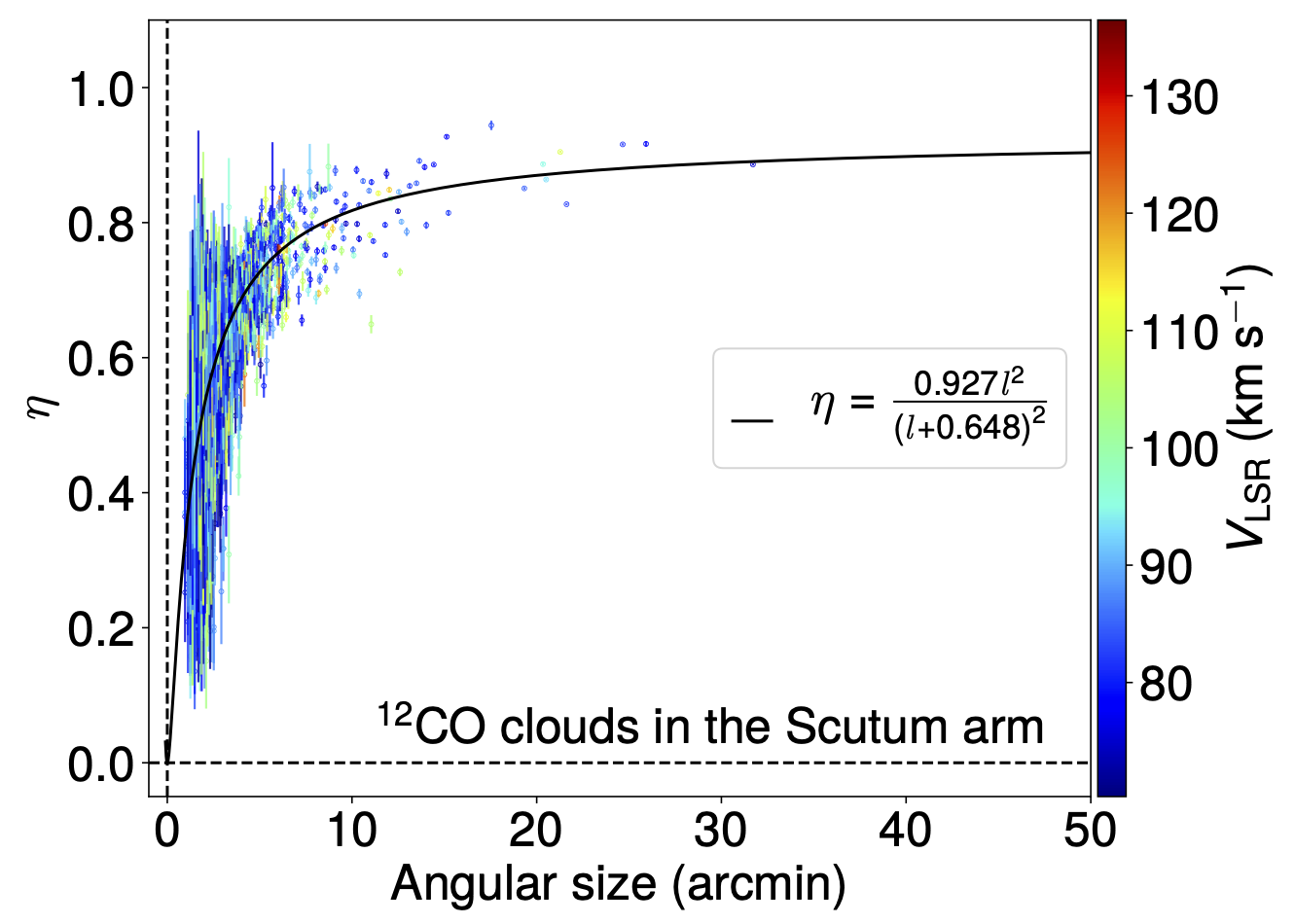}{0.46\textwidth}{(c)}  
   \fig{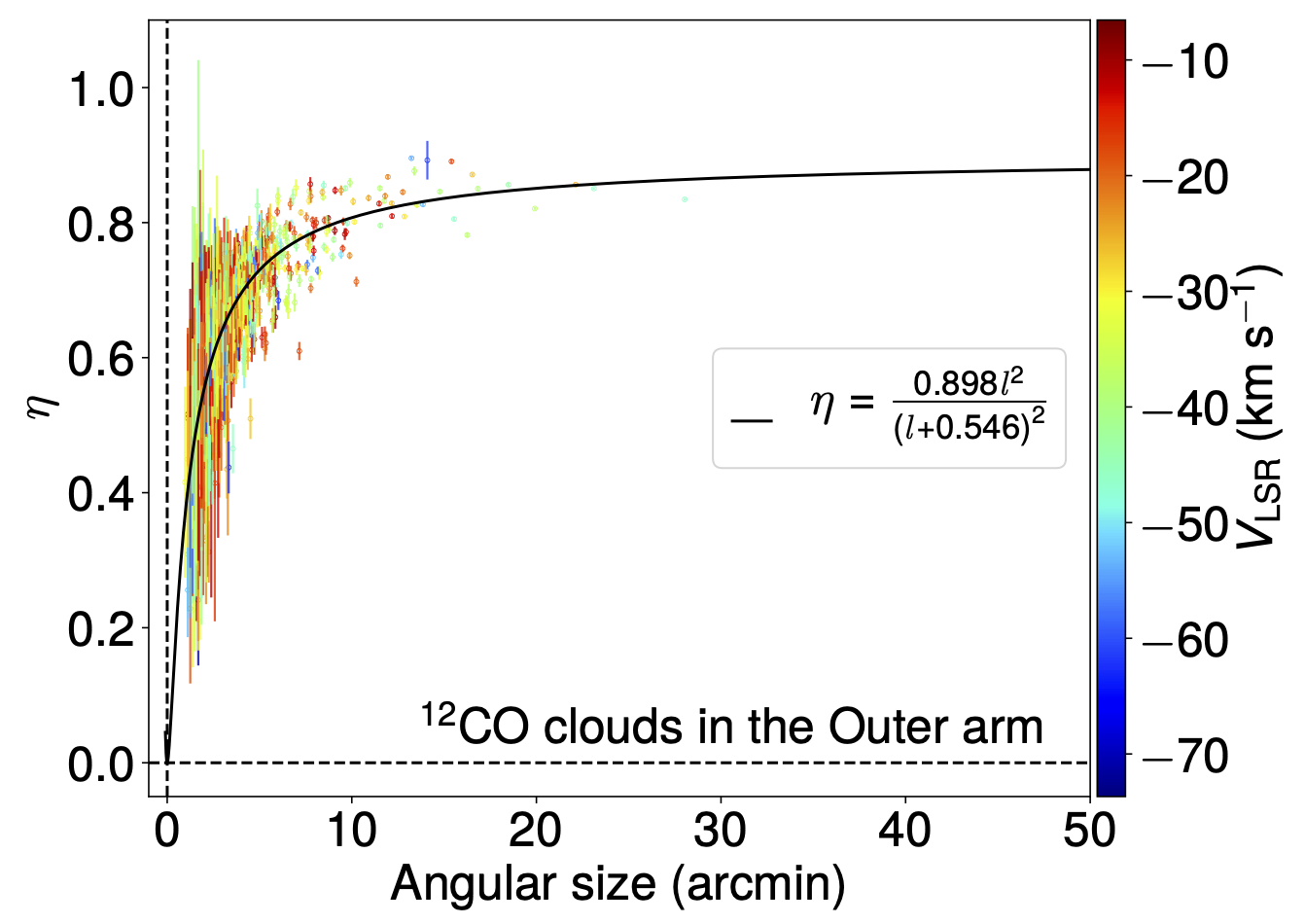}{0.46\textwidth}{(d) } 
 } 
\caption{$\eta$ of individual \cofs\ clouds against their angular sizes in four spiral arm segments: (a) the Local arm, (b) the Sagittarius arm, (c) the Scutum arm, and (d) the Outer arm. The color code represents $V_{\rm LSR}$, and see Equation \ref{eq:ffsize} for the form of black  solid lines.  \label{fig:sizebffco12} } 
\end{figure}

\subsection{Sample-based beam filling factors}


The procedure of deriving beam filling factors for individual molecular clouds is similar to that of image-based beam filling factors,  and the only difference is that the mean $T_{\rm mb}$ for each molecular cloud is calculated over its own region. This region is determined with raw (unsmoothed) MWISP data using DBSCAN (down to 2$\sigma$), and voxels involved in the estimate of the mean $T_{\rm mb}$ are required to be above 2$\sigma$ level in all smoothing cases.

The beam filling factor of each molecular cloud is estimated with Equation \ref{eq:fffunction} based on the variation of mean $T_{\rm mb}$ with respect to the beam size. As examples, we show $\eta$ of four \cofs\ molecular clouds in the Local arm in Figure \ref{fig:bfffour}. Usually, molecular clouds whose mean $T_{\rm mb}$ decreases approximately linearly have high beam filling factors, while molecular clouds with exponentially decreasing mean $T_{\rm mb}$ have low beam filling factors.

We found that $\eta$ is correlated with the angular size $l$  of molecular clouds. The angular size is defined as an equivalent diameter derived with 
\begin{equation}
l= \sqrt{\frac{4A}{\pi}-\Theta_{\rm MWISP}^2},
\label{eq:size}
\end{equation}
where $A$ is the angular area and $\Theta_{\rm MWISP}$ is the beam size of the MWISP survey. Figure \ref{fig:sizebffco12} demonstrates the $\eta$ variation of \cofs\ clouds against their angular sizes. Evidently, compared with the radial velocity (the color code), which is usually used as a distance indicator, $\eta$ is more related to the angular size. $\eta$ is approximately unity for molecular clouds with large angular sizes ($\geq$ 10\arcmin), but decrease sharply for small ones.

Given the large dispersion of $\eta$, we only look for a first-order approximation for the relationship between $\eta$ and $l$. The function we used is  
\begin{equation}
\eta= \eta_{\rm max} \frac{l^2}{\left(l+l_{1/4}\right)^2},
\label{eq:ffsize}
\end{equation} 
where $l_{1/4}$ is the angular size corresponding to $\eta=0.25\eta_{\rm max}$. With this model, the range of $\eta$ is from 0 ($l\rightarrow0$) to  $\eta_{\rm max}$ ($l\rightarrow\infty$),  and compared with other models (see Section \ref{sec:disbffsize}), Equation \ref{eq:ffsize} yields a smaller rms residual and has a clear physical meaning. Theoretically, $\eta_{\rm max}$ should equal one, but $\eta_{\rm max}$ is slightly less than one in practice, possibly due to the error of simulated data. The right side of Equation \ref{eq:ffsize} is a ratio of the angular area of molecular clouds to the observed angular area enlarged by the beam, consistent with the beam filling factor definition. Consequently, we use Equation \ref{eq:ffsize} to model the relationship between $\eta$ and $l$. $\eta_{\rm max}$ and $l_{1/4}$ is solved with \texttt{curve\_fit} of the \texttt{Python} package \texttt{SciPy}, considering the error of $\eta$. 

Results of \cofs\ molecular clouds are described in Figure \ref{fig:sizebffco12}, and values of $l_{1/4}$ for three CO lines and four spiral arms are summarized in Table \ref{tab:cloudnumber}.  Remarkably, values of  $\eta_{\rm max}$ and  $l_{1/4}$ is approximately equal for molecular clouds in all four arms, suggesting that molecular clouds with close angular sizes have approximately equal beam filling factors despite being at different distances.   
 


\begin{figure}[ht!]
\plotone{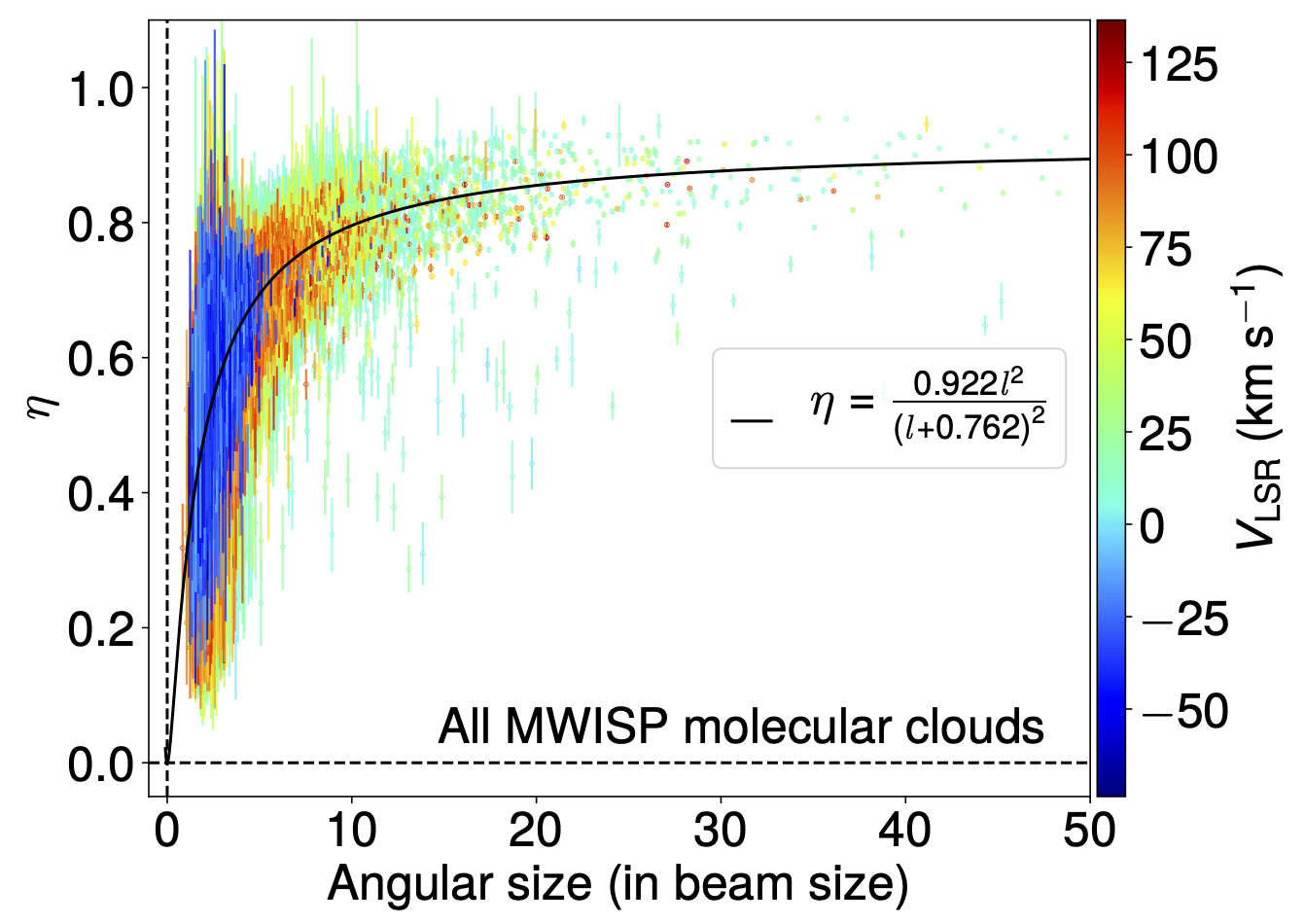}
\caption{$\eta$-$l$ relationship with all molecular cloud samples in four spiral arm segments,  including \cofs, \coss, and \cots\ samples. See Equation \ref{eq:overallBFF} for the form of the black solid line.  \label{fig:bffsizeall} }
\end{figure}

 Given the similarity of $\eta_{\rm max}$ and $l_{1/4}$ in four arm segments, we fit overall values with all molecular clouds. As demonstrated in Figure \ref{fig:bffsizeall}, the overall fitting gives values of $l_{1/4} = 0.762\pm0.001$ and $\eta_{\rm max}=0.922\pm0.000$ , i.e., $\eta$ is approximately 
\begin{equation}
\eta = \frac{0.922l^2}{\left(l+0.762\right)^2},
\label{eq:overallBFF}
\end{equation}
where $l$ is the angular size of molecular clouds in units of the beam size. 

\begin{deluxetable*}{cccccc} 
\tabletypesize{\scriptsize}

\tablecaption{Beam filling and sensitivity clip factor relationships in the first Galactic quadrant based on the MWISP survey. \label{tab:cloudnumber}}
\tablehead{
\colhead{} & \colhead{}&     \multicolumn{2}{c}{$\eta=\eta_{\rm max} \frac{l^2}{\left(l+l_{1/4}\right)^2}$}    &   \multicolumn{2}{c}{$\xi=\frac{\left(x-x_0\right)^2}{\left(x-x_0+x_{1/4}\right)^2}$} \\
  \cline{3-4} \cline{5-6} 
\colhead{Arm} & \colhead{Line}&       \colhead{$\eta_{\rm max}$} & \colhead{$l_{1/4}$} &   \colhead{$x_0$} &  \colhead{$x_{1/4}$}  
} 
\colnumbers 
\startdata  
& \cofs & 0.930 $\pm$ 0.000 & 0.617 $\pm$ 0.002 & 2.228 $\pm$ 0.001 & 0.455 $\pm$ 0.000   \\
Local & \coss & 0.913 $\pm$ 0.000 & 0.593 $\pm$ 0.002 & 2.224 $\pm$ 0.001 & 0.456 $\pm$ 0.000   \\
 & \cots & 0.896 $\pm$ 0.001 & 0.474 $\pm$ 0.009 & 2.065 $\pm$ 0.005 & 0.523 $\pm$ 0.002   \\
\hline
 & \cofs & 0.939 $\pm$ 0.000 & 0.694 $\pm$ 0.002 & 2.215 $\pm$ 0.001 & 0.460 $\pm$ 0.000   \\
Sagittarius & \coss & 0.911 $\pm$ 0.000 & 0.628 $\pm$ 0.002 & 2.228 $\pm$ 0.001 & 0.463 $\pm$ 0.000   \\
 & \cots & 0.889 $\pm$ 0.002 & 0.542 $\pm$ 0.007 & 2.187 $\pm$ 0.004 & 0.472 $\pm$ 0.002   \\
\hline
 & \cofs & 0.927 $\pm$ 0.001 & 0.648 $\pm$ 0.002 & 2.236 $\pm$ 0.001 & 0.448 $\pm$ 0.000   \\
Scutum & \coss & 0.916 $\pm$ 0.001 & 0.657 $\pm$ 0.003 & 2.223 $\pm$ 0.002 & 0.451 $\pm$ 0.001   \\
 & \cots & 0.864 $\pm$ 0.001 & 0.503 $\pm$ 0.005 & 2.204 $\pm$ 0.004 & 0.468 $\pm$ 0.001   \\
\hline
 & \cofs & 0.898 $\pm$ 0.001 & 0.546 $\pm$ 0.004 & 2.248 $\pm$ 0.002 & 0.451 $\pm$ 0.001   \\
Outer & \coss & 0.864 $\pm$ 0.005 & 0.538 $\pm$ 0.017 & 2.138 $\pm$ 0.010 & 0.503 $\pm$ 0.004   \\
   & \cots & -- & -- & -- & --   \\ 
\enddata
\tablecomments{  $l_{1/4}$ is in arcmin, while $x_0$ and  $x_{1/4}$ are in rms noise.}
\end{deluxetable*}

 \begin{figure}[ht!]
 \gridline{\fig{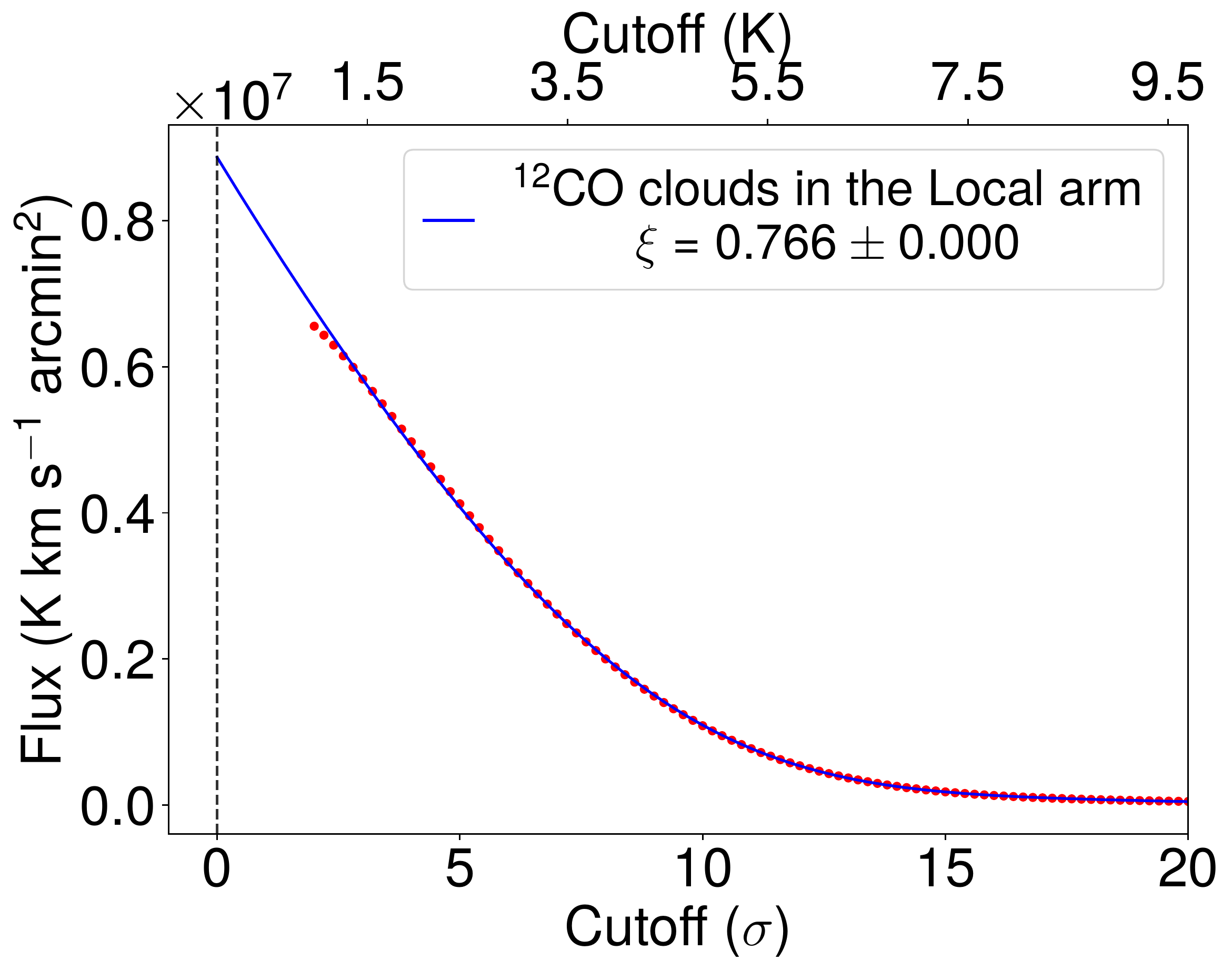}{0.33\textwidth}{(a)}  \fig{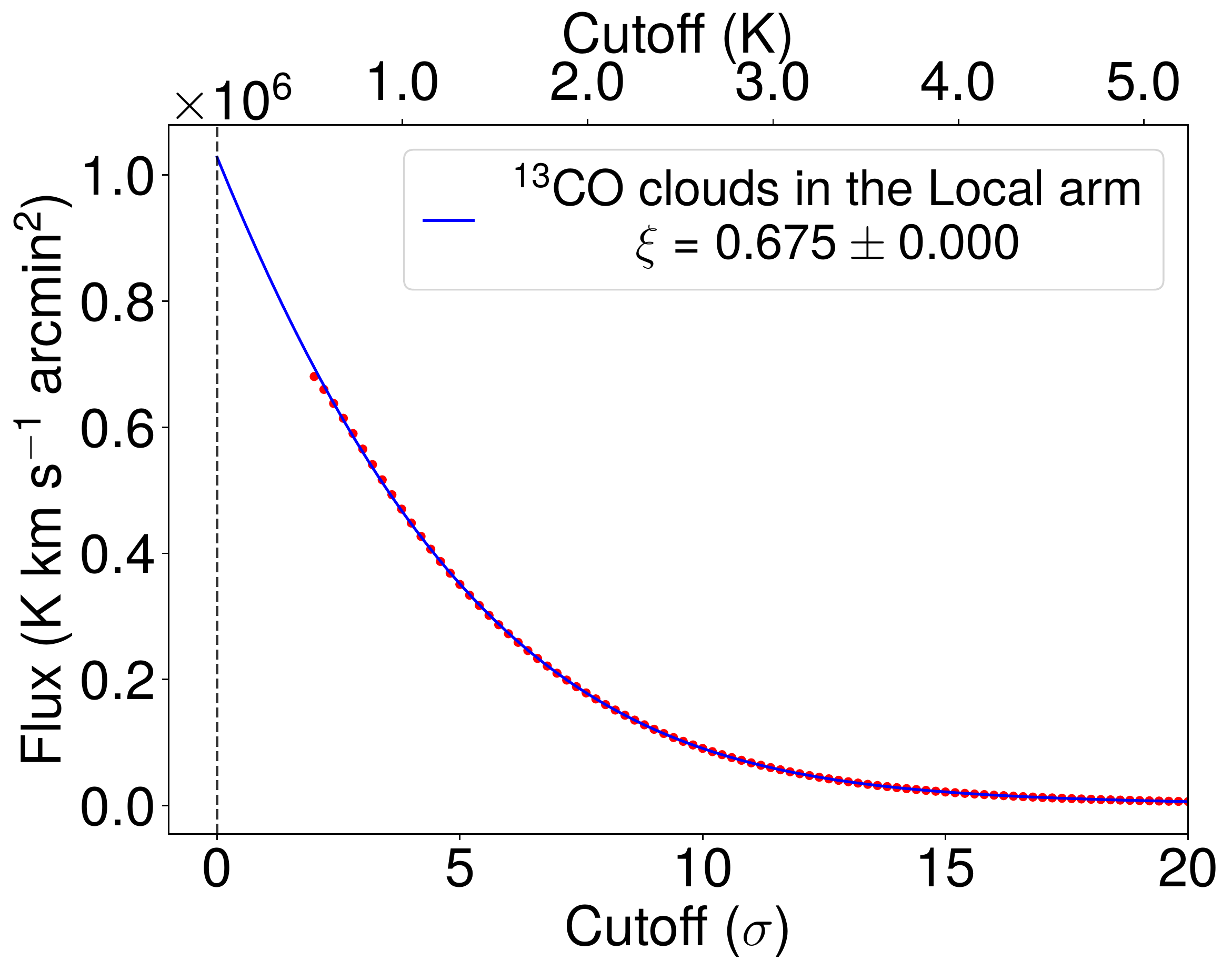}{0.33\textwidth}{(b)}   \fig{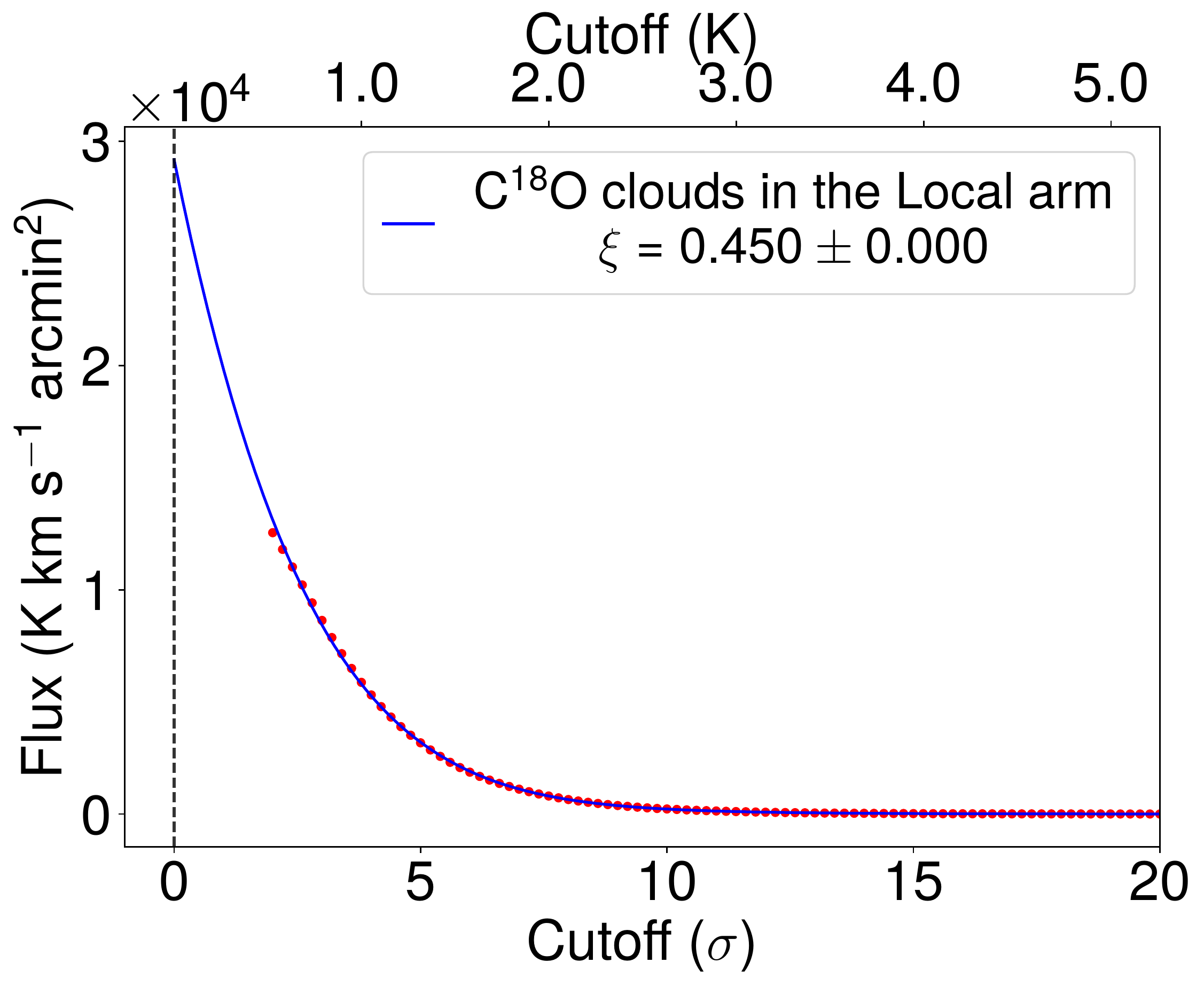}{0.33\textwidth}{(c)} 
 } 
\caption{ Fitting of image-based flux variation with respect to cutoffs in the Local arm: (a) \cofs, (b) \coss, and (c) \cots. Flux variations (blue lines) are modeled with Equation \ref{eq:modelsens}, and the corresponding $\xi$ (see Equation \ref{eq:ffmwispcut} for the definition) is derived with the ratio of modeled flux values at 2$\sigma$ to that at zero. The error bar is smaller than the marker size. \label{fig:fflocalsen} } 
\end{figure}

 \begin{figure}[ht!]
 \gridline{\fig{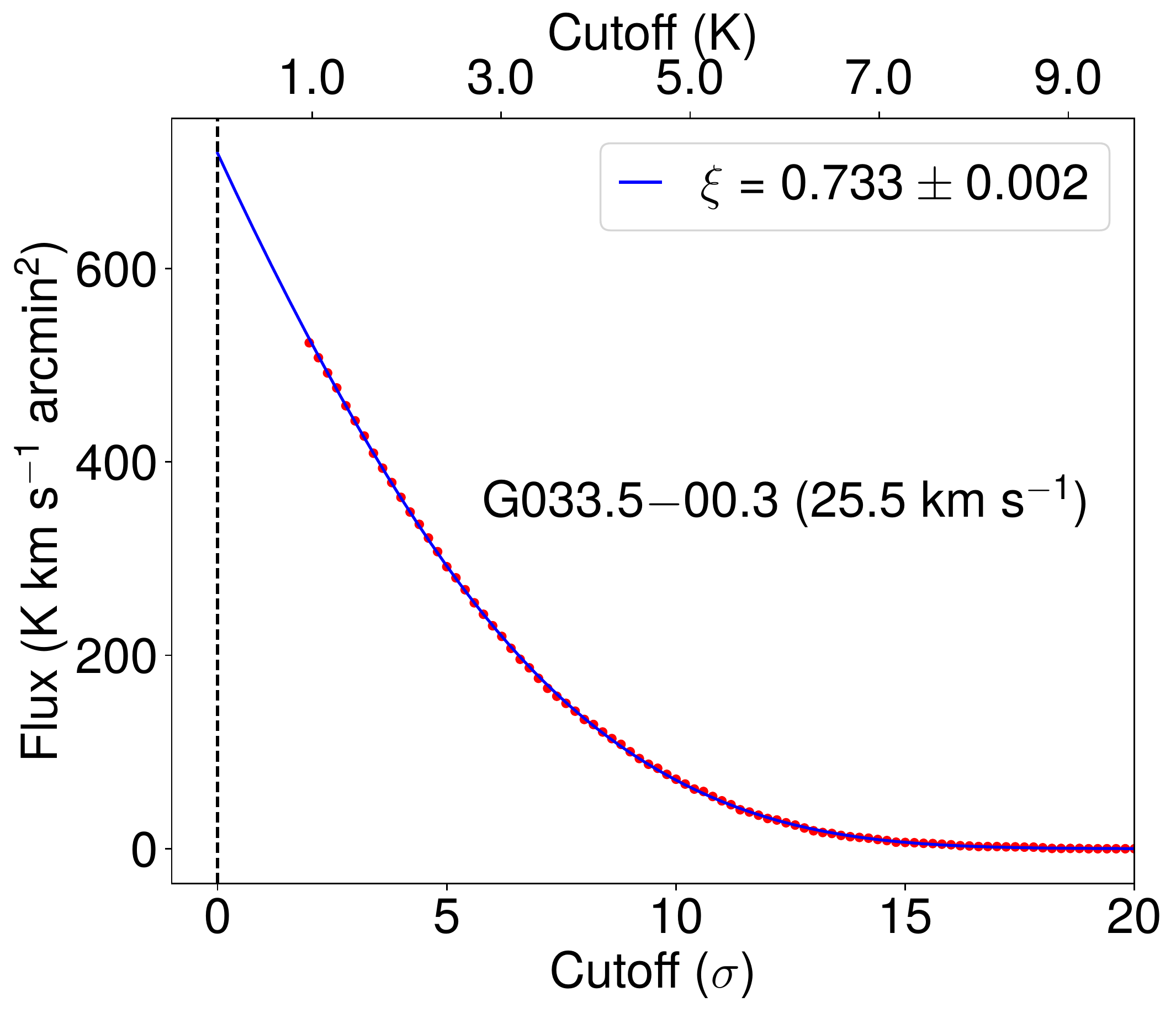}{0.33\textwidth}{(a)}  \fig{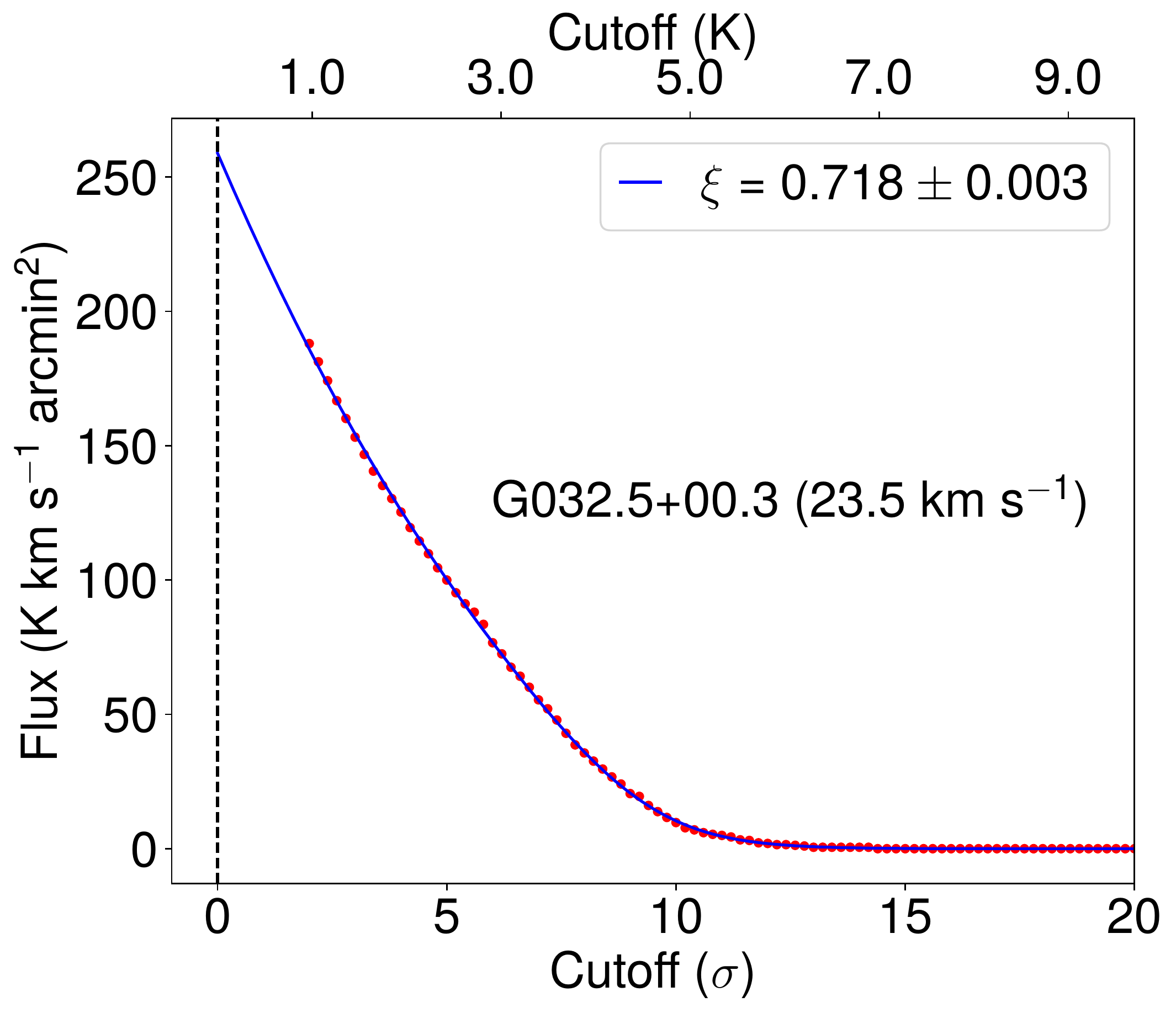}{0.33\textwidth}{ (b)}   \fig{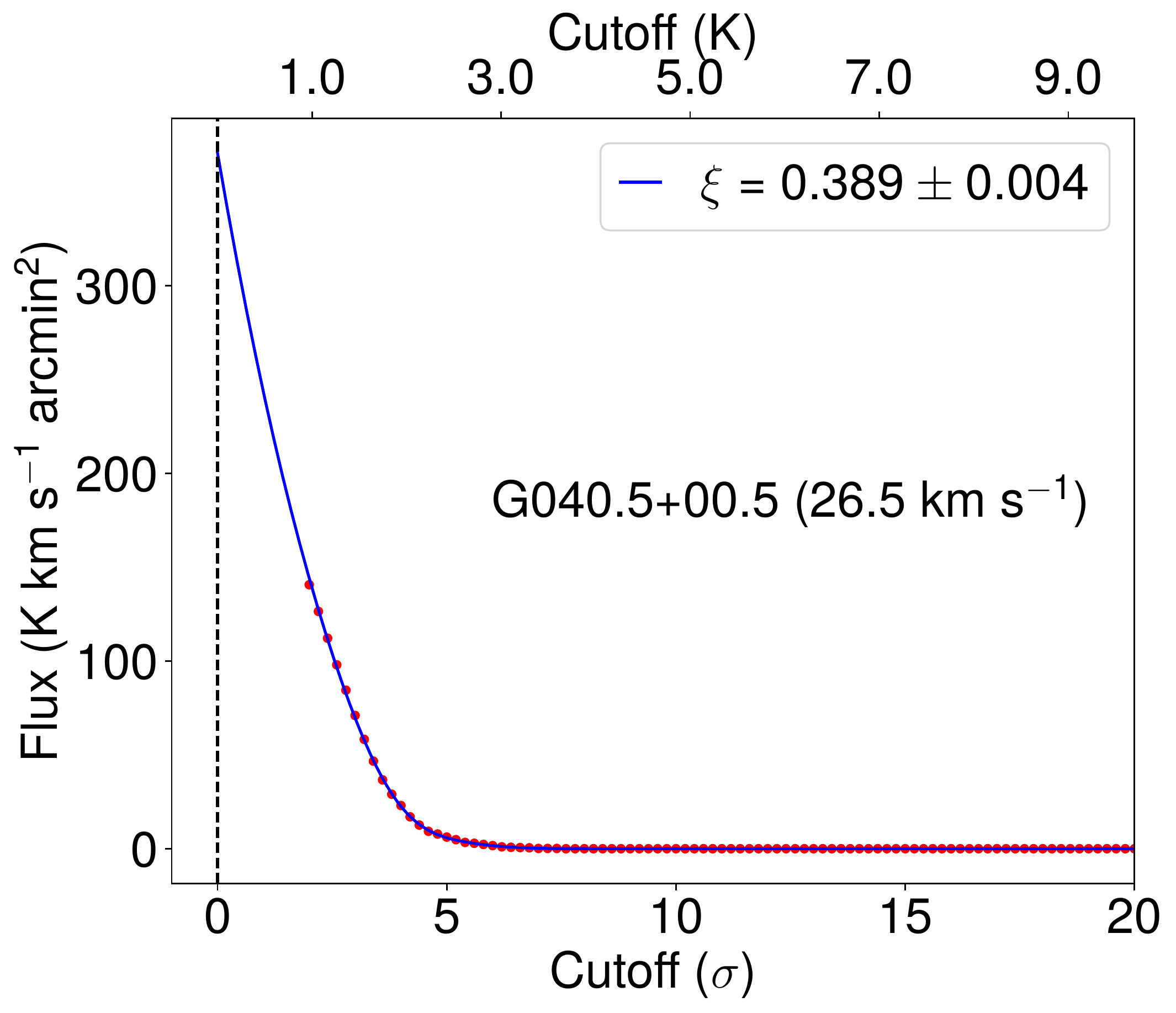}{0.33\textwidth}{(c)} 
 } 
\caption{ Same as Figure \ref{fig:fflocalsen} but for three \cofs\ clouds in the Local arm: (a) G033.5$-$00.3 at 25.5 \kms, (b) G032.5$+$00.3 at 23.5 \kms, and (c) G040.5$+$00.5 at 26.5 \kms.    \label{fig:ffsensthree} } 
\end{figure}

\begin{figure}[ht!]
\plotone{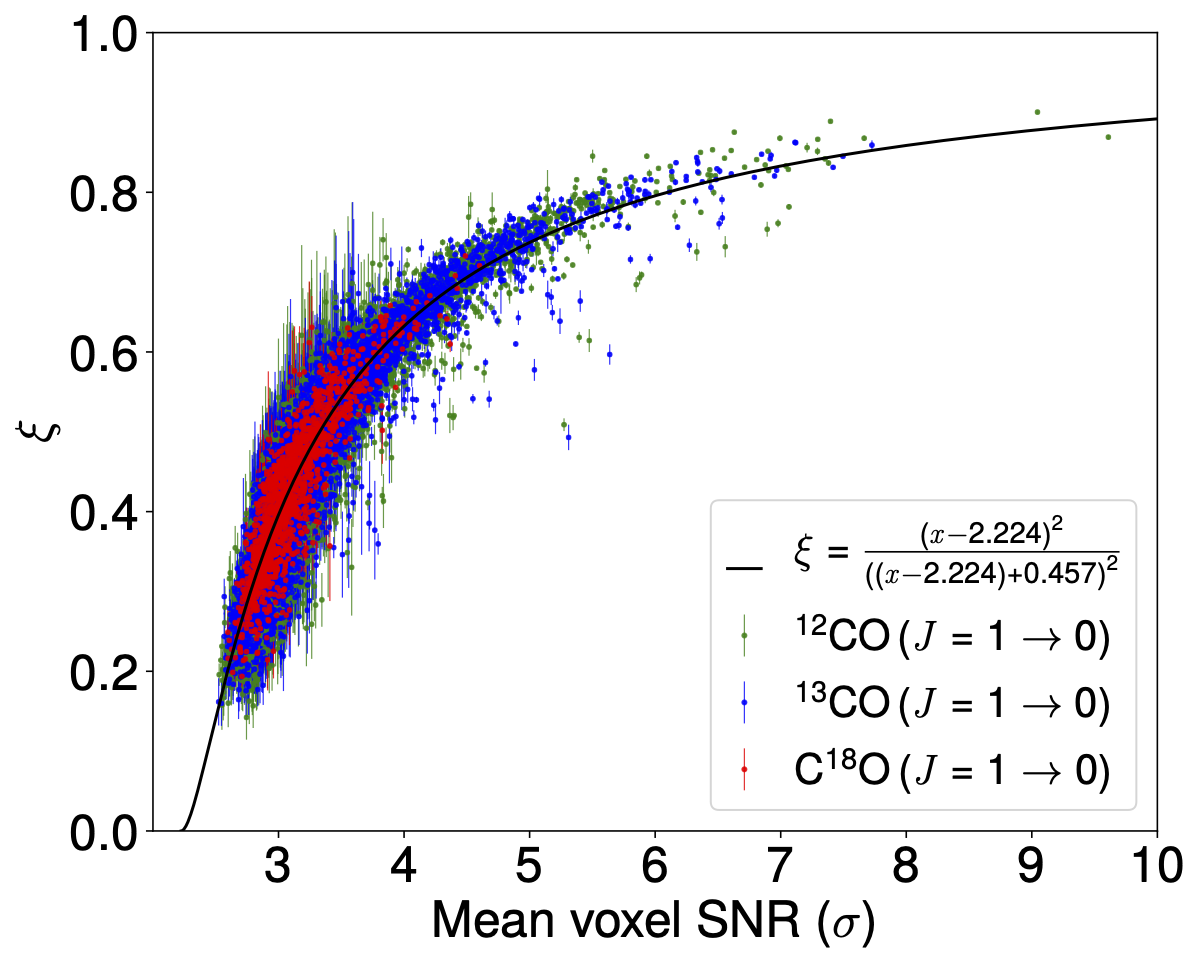}
\caption{ The relationship between the sample-based $\xi$ and the mean voxel SNR. The mean voxel SNR is the mean SNR of all voxels in a molecular cloud. In the relationship fitting, all MWISP molecular clouds samples are used, including four arm segments and three CO isotopologue lines. For clarity purposes, only molecular clouds with relative errors less than 20\% are displayed. The mean voxel SNR is in units of rms noise, and see Equation \ref{eq:modelsens} for the black solid line.  \label{fig:ffcompare} }
\end{figure}

\subsection{Image-based sensitivity clip factors}

Similar to the beam filling factor, the sensitivity clip factor can also be estimated by taking all emission as a single object. Figure \ref{fig:fflocalsen} displays results of $\xi$ for local  molecular clouds.  As can be seen, Equation \ref{eq:modelsens} fits the flux variation well, but show slight deviations  around 2$\sigma$ cutoff.  This is because at  2$\sigma$, the observed flux may not be complete  due to the insufficient SNR.  

Table \ref{tab:mwispfilling} lists results of all four arm segments. Clearly, among three CO lines, \cofs\ has the highest $\xi$, while \cots\ has the smallest $\xi$.  As to arm segments, the Scutum and Outer arm has the highest and lowest $\xi$, respectively, while the rest two arms have medium $\xi$.

\subsection{Sample-based sensitivity clip factors}

 To make $\xi$  consistent with $\eta$, the minimum cutoff of brightness temperature for individual molecular clouds is 2$\sigma$. In Figure \ref{fig:ffsensthree}, fitting results  show that  Equation \ref{eq:modelsens} describes the flux variation well for individual molecular clouds.

We found that $\xi$ are correlated with the mean voxel SNR of molecular clouds.  This relationship is insensitive to molecular cloud tracers and distances,  and can be described with Equation \ref{eq:ffsize} but with a slight adjustment of zero points:
 \begin{equation}
\xi=\frac{\left(x-x_0\right)^2}{\left(x-x_0+x_{1/4}\right)^2},
\label{eq:ffsnr}
\end{equation}
where $x_0$ is the zero point and  $x_{1/4}$ is the mean voxel SNR (with respect to $x_0$) at which $\xi=1/4$.

 Figure \ref{fig:ffcompare} demonstrates the relationship between the sensitivity clip factor and the mean voxel SNR, including molecular cloud samples of all four arm segments and three CO lines. The molecular cloud samples only contain normal fittings (25542 in total) of Equation \ref{eq:modelsens}, i.e., $a$, $b$, and $\delta$ are  positive and $0<\xi<1$.


 

 \section{Discussion}
 \label{sec:discuss}
 \subsection{Simulated data} 
  \label{sec:smdata} 
 In this work, we used simulated data instead of practical observations, which may cause systematic errors. The raw Data is clipped at a certain sensitivity level, and all smoothing cases are based on clipped images. Consequently, $T_{\rm mb}$ in simulated data is possibly systematically smaller (than practical observations) due to the clip effect of the raw data, particularly for voxels near the edge of molecular clouds. 
 
 This systematic shift of simulated $T_{\rm mb}$ is demonstrated in Figure 1. The mean $T_{\rm mb}$ of the raw data is slightly larger than the fitted value. This discrepancy can be examined with practical observations, which do not have this issue.

\subsection{Beam filling factors and the angular size}
\label{sec:disbffsize}

Although we use Equation \ref{eq:ffsize} to describe the relationship between the beam filling factor and the angular size, the choice of functions is not unique. We compared two additional function forms, and found that judging by the rms residual, Equation \ref{eq:ffsize} outperforms the other two models. One of the two models uses the function   
\begin{equation}
\eta = \frac{l^2}{\left(l^2+a\right)},
\label{eq:ffsize1}
\end{equation}
where $a$ is a parameter. Equation \ref{eq:ffsize1} resembles the convolution of Gaussian distributions \citep{2008A&A...482..197P}, while the third model has a form of 
\begin{equation}
\eta=a(1-\exp(-bl)),
\label{eq:ffsize2}
\end{equation}
where $a$ and $b$ are two parameters and $l$ is the angular size of molecular clouds. In this case, $\eta\rightarrow0$ as $l\rightarrow0$, while  $f\rightarrow a$ as $l\rightarrow\infty$, i.e., $a$ is the maximum beam filling factor.

To test which model performs best, we split \cofs\ molecular cloud samples in the Local arm into two categories: (1) the training set and (2) the validation data set. The training set is used to fit the model, while the validation data is used to verify the model.  We examined two cases, having 20\% and 30\% validation data ratios, respectively, and the weighted rms residual (chi-square) of the validation data is used as an indicator of modeling qualities. As shown in Figure \ref{fig:testmodel},  Equation \ref{eq:ffsize} possess the best performance.




 \begin{figure}[ht!]
 \gridline{
 \fig{{modelTestTestRatio0.20}.png}{0.48\textwidth}{(a) } 
 \fig{{modelTestTestRatio0.30}.png}{0.48\textwidth}{(b)}  
 } 
\caption{A comparison of three functions to model the relationship of $\eta$ and the angular size ($l$). The molecular cloud samples used are \cofs\ clouds in the Local arm.  Panel (a) uses 80\% of the samples to fit the model and the rest 20\% for testing, while the validation data ratio in panel (b) is 30\%. The right side of each panel plots the variation of the weighted rms residual of the validation samples with respect to a maximum relative error threshold for each model. See Equation \ref{eq:ffsize} (blue), \ref{eq:ffsize1} (green), and \ref{eq:ffsize2} (red) for the form of three models. As examples, on the left side of each panel, we display the training (gray) and validation (black) samples (maximum relative errors less than 0.2), together with the fitting of three models. \label{fig:testmodel} } 
\end{figure}

\begin{figure}[ht!]
\plotone{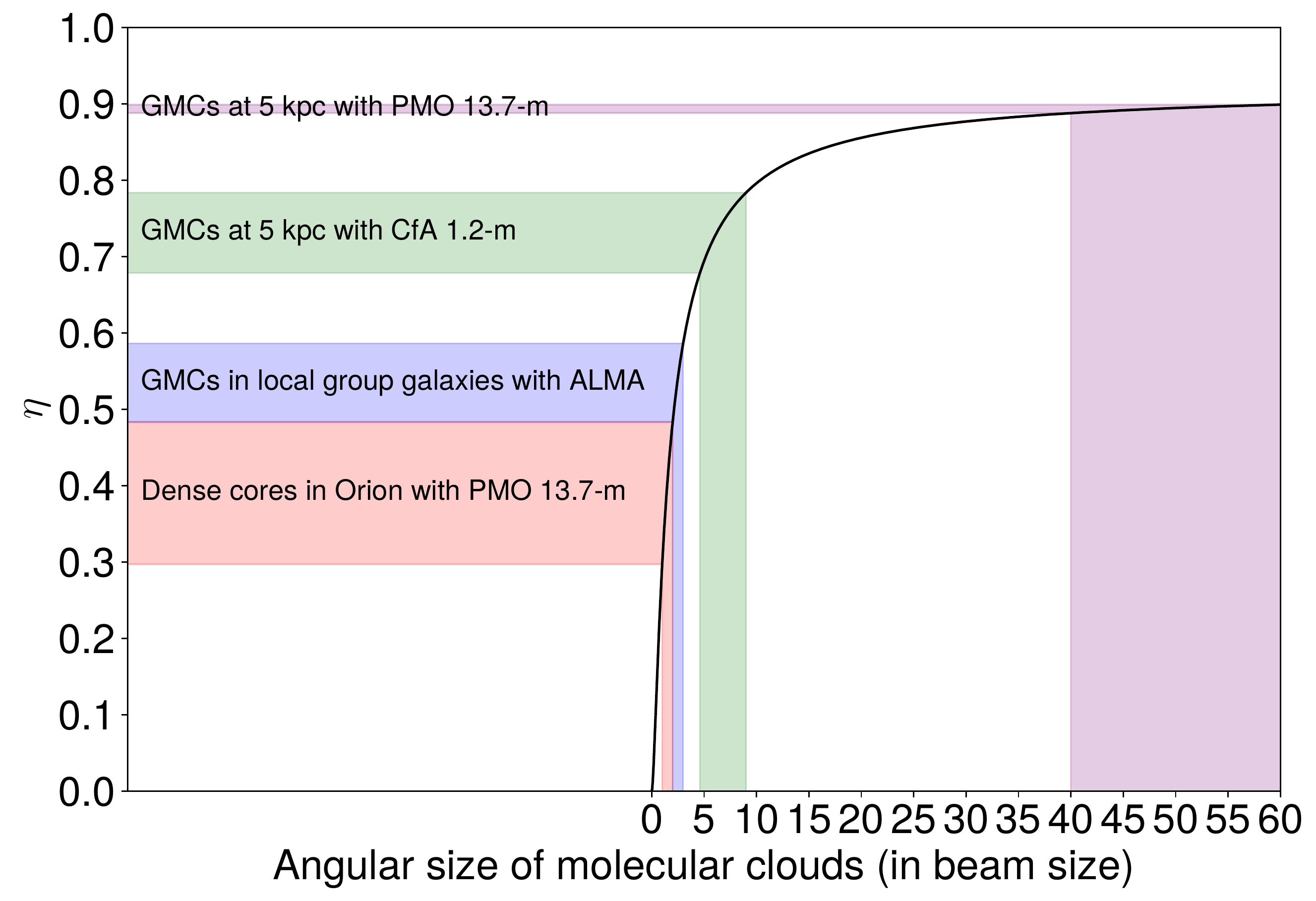}
\caption{Beam filling factors of molecular clouds in different observational cases. See Equation \ref{eq:overallBFF} for the form of the black solid line. We list four typical situations with respect to GMCs (two Galactic cases and an extragalactic one) and dense cores (in Orion). \label{fig:bffconcept} }
\end{figure}

\subsection{Beam filling factors of molecular clouds}

 Beam filling factors of small molecular clouds are largely uncertain, while   beam filling factors of large molecular clouds are well modeled. According to the relationship between the beam filling factor and the angular size of molecular clouds, beam filling factors are approximately unity for relatively large molecular clouds, and  decrease fast toward small molecular clouds. Beam filling factors are less than 0.5 for molecular clouds with angular size less than $\sim$2 beam sizes (after deconvolution), and given the large uncertainty, it cloud be even smaller. 

 Molecular cloud samples in this work is built with the DBSCAN detection scheme, but an alternative algorithm would yield different molecular cloud samples. The variation of beam filling and sensitivity clip factors with respect to molecular cloud samples is possibly significant and will be investigated in the future. 

 Due to the uncertainty of beam filling factors, estimations of excitation temperatures and optical depths for small molecular clouds are subject to large errors. This is usually the case for extragalactic observations, in which most molecular clouds are unresolved. At least a factor of 2 should be used to calibrate the brightness temperature in the application of radiative transfer equations.

We estimate beam filling factors of observations toward Giant molecular clouds (GMCs) in size of $\sim$50-100 pc \citep{2018ARA&A..56...41M} and dense cores in size of $\sim$0.1-0.2 pc \citep{2018ARA&A..56...41M}  based on Equation \ref{eq:overallBFF}. As demonstrated in Figure \ref{fig:bffconcept}, GMC observations with ALMA toward the Local Group galaxies \citep{2018ApJ...860..172S} have an angular size of 2-3 beam sizes, and the corresponding $\eta$ is about 0.5. For those observations, the $\alpha_{\rm vir}$ would be overestimated by a factor of 2. For GMCs at a medium distance of $\sim$5 kpc in the Milky Way, observations of the CfA 1.2-m \citep{2001ApJ...547..792D}  yield  $\eta$ of $\sim$ 0.75, and with PMO 13.7-m \citep{2019ApJS..240....9S}, the $\eta$ of GMCs would be approximately 0.9. Seen by PMO 13.7-m, dense cores in a close high-mass star forming region, Orion ($\sim$400 pc) have an angular size of 1-2 beam sizes, and their $\eta$ would be about 0.4. Consequently, observations with relatively low angular resolutions would significantly underestimate the brightness temperature.

 \begin{deluxetable}{ccccccccc}
\tablecaption{Observation parameters of four surveys.\label{Tab:surveyParameters}}
\tablehead{
\colhead{Survey Name} & \colhead{Spectral line} & \colhead{Beam size} & \colhead{$l$ } &\colhead{$b$} & \colhead{$V_{\rm LSR}$} & \colhead{$\delta v$}  &  \colhead{ rms noise }  & \colhead{Cloud number} \\ 
\colhead{ } & \colhead{} &  &\colhead{([\deg, \deg])} &\colhead{([\deg, \deg])} & \colhead{ (\kms) } & \colhead{ (\kms) } & \colhead{(K)} &
  \colhead{ } 
}

\startdata
CfA 1.2-m\tablenotemark{a}  &  \cof\ & 8.5\arcmin  & [17, 75]  &  [-3, 4.9] & [-87, 140] & 1.3& $\sim$0.1   &    243 \\\
GRS\tablenotemark{b} &   \cos\   &  46\arcsec  &[25.8, 49.7] & [-1, 1]    & [-5, 70]  &0.2 & $\sim$0.1   &     6721 \\
OGS\tablenotemark{c} &   \cof\ &   46\arcsec &[102.5, 141.5]    & [-3, 5.4]  &   [-100, 20]&0.8  & $\sim$0.6 &    4928 \\
COHRS\tablenotemark{d} &    \cotss  &   16\arcsec & [24.75, 48.75]     & [-0.275, 0.275]   & [-5, 70] & 1.0  & $\sim$0.4 &   11877 \\
\enddata  
\tablenotetext{}{($a$) \citet{2001ApJ...547..792D}. ($b$) \citet{2006ApJS..163..145J}. ($c$) \citet{1998ApJS..115..241H}. ($d$) \citet{2013ApJS..209....8D}.}  
\end{deluxetable}

 \begin{figure}[ht!]
 \gridline{ \fig{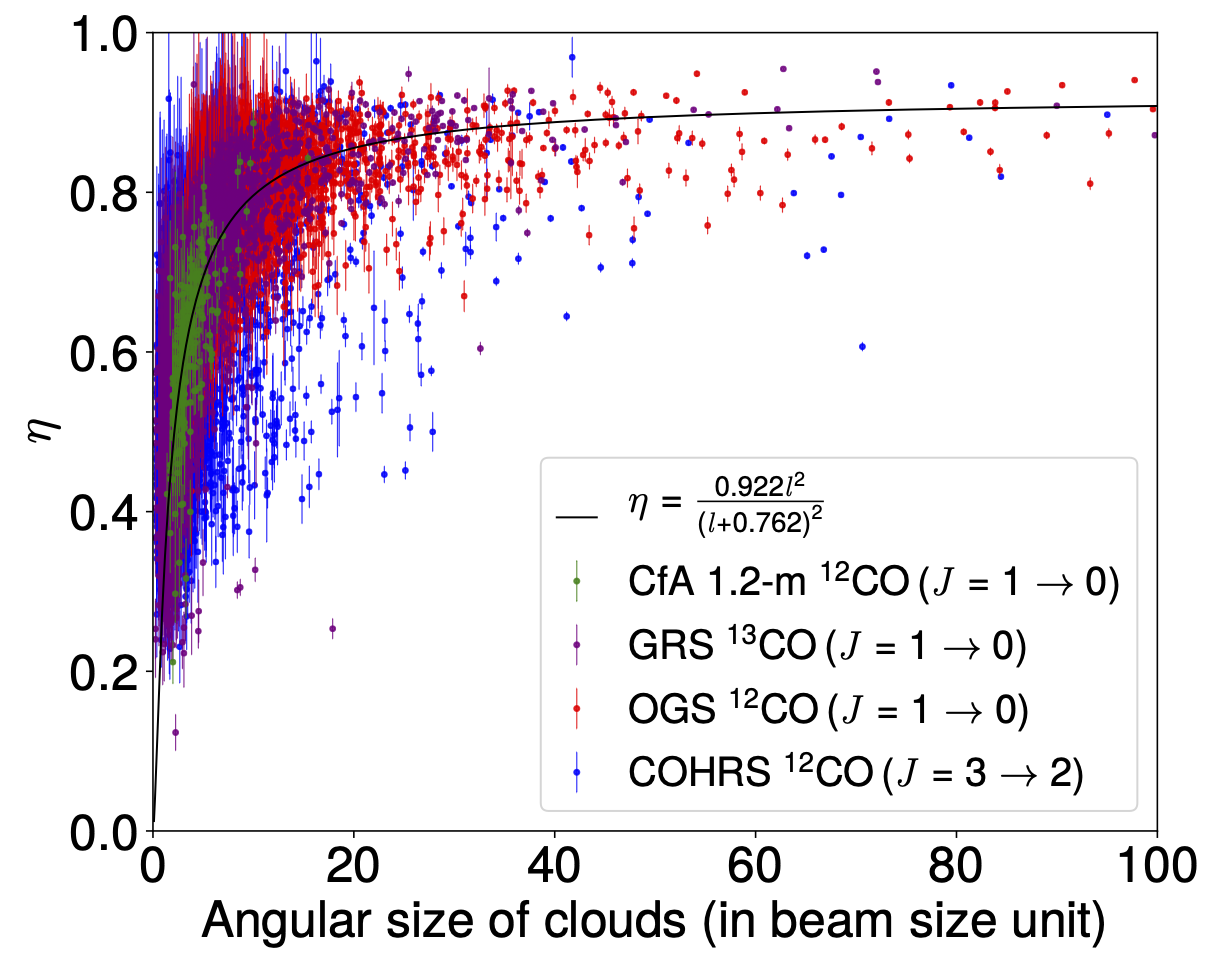}{0.45\textwidth}{(a)  }   \fig{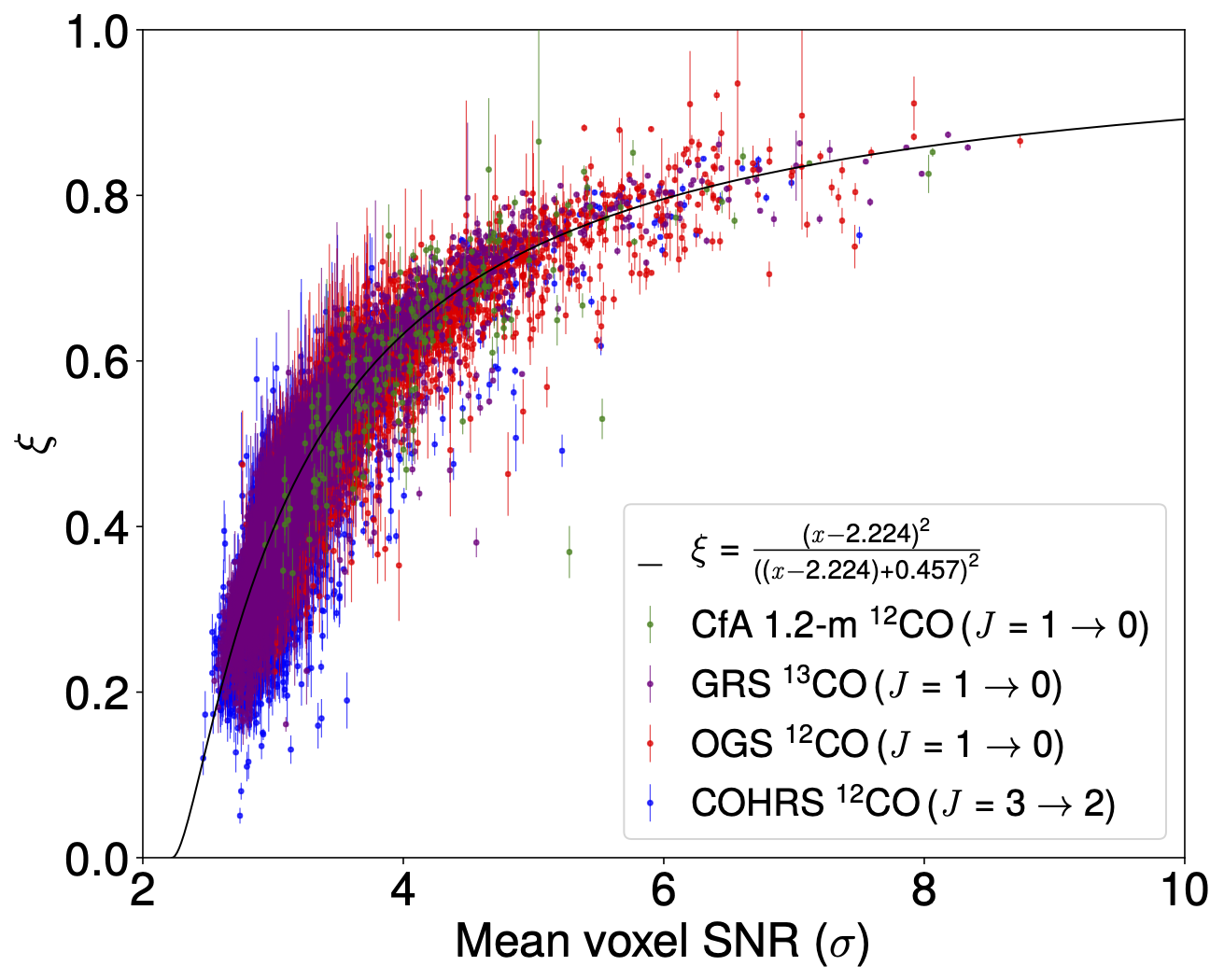}{0.45\textwidth}{(b)  } 
 } 
\caption{Beam filling and sensitivity clip factors of molecular clouds in four CO surveys (see Table \ref{Tab:surveyParameters}). For clarity purposes, molecular clouds with beam filling factor relative errors larger than 20\% were removed. As a comparison, we display results derived with the MWISP survey in black solid lines: (a) the  $\eta$ and the angular size relationship (see Equation \ref{eq:overallBFF}) and (b) the $\xi$ and the mean voxel SNR relationship (see Figure  \ref{fig:ffcompare}). \label{fig:bffsurvey}  } 
\end{figure}


 \subsection{Sensitivity clip factors of molecular clouds}

 Results of the sensitivity clip factors reveal that measured flux is incomplete.  According to the relationship in Figure \ref{fig:ffcompare}, the sensitivity clip factor is about 0.5 for a cloud with a mean voxel SNR of 3.3. This suggests that a large fraction of flux is missed for barely detected molecular clouds. Apparently, physical properties of \cot\ are  largely uncertain due to their low beam filling and sensitivity clip factors.

 The sensitivity clip factor is remarkably consistent between molecular clouds. This universality suggests that the brightness temperature distribution of most molecular clouds is similar.  As shown in Figure \ref{fig:ffcompare}, the dispersion of $\xi$ is small, meaning that molecular clouds with the same mean voxel SNR miss a similar fraction of flux, this only happens when their distributions of brightness temperatures are the same.

\subsection{Comparison with other surveys}

In order to see the variation of beam filling sensitivity clip factors under other observations with different spectral line tracers, beam sizes, and sensitivities, we compare four CO surveys with the MWISP survey. Table \ref{Tab:surveyParameters} lists observational parameters and PPV ranges of four examined surveys, and for the large-scale \cofs\ survey \citep{2001ApJ...547..792D} conducted with CfA 1.2-m, we choose a uniformly sampled region that has a large Galactic latitude coverage in the first Galactic quadrant.

With the same smoothing and cloud identification procedure, we calculated $\eta$ and $\xi$ of molecular clouds for the four CO surveys. As demonstrated in Figure \ref{fig:bffsurvey}, their beam filling and sensitivity clip factors are remarkably consistent with that derived with the MWISP survey. This indicates that despite of having different sensitivities, beam sizes, and even spectral lines, beam filling and sensitivity clip factors of molecular clouds show similar relationships with molecular cloud sizes and mean voxel SNRs.

Our analysis methodology is applicable to other phases of the ISM. \citet{1999ptep.proc...61B} studied the volume filling factor of multiple phases of the ISM, including \HII, \HI,  molecular, and dust clouds, and the results suggest similar structures for ionized and molecular clouds. It is interesting to compare the results with  directly measured beam filling and sensitivity clip  factors of the ISM, which could possibly reveal the structure and the distribution of the ISM.




\section{Summary}
\label{sec:summary}

We studied beam filling and sensitivity clip factors of molecular clouds by simulating observations with large beam sizes and low sensitivities using the MWISP CO survey in the first Galactic quadrant.  The beam filling factor is used to calibrate the brightness temperature, and the sensitivity clip factor is used to  estimate the completeness of the flux. The beam filling factor is modeled with a two-component function according to the variation of the mean $T_{\rm mb}$ with respect to beam sizes,  while the sensitivity clip factor is modeled using a quadratic function with a fast decreasing tail. In order to examine the collective and individual properties, we derived  beam filling and sensitivity clip factors  based on both the entire images and molecular cloud samples.

 The main results can be summarized as follows:
 \begin{enumerate}
\item   Beam filling factors  of \cofs\ and \coss\ are approximately unity in the Local ($\sim$1 kpc), the Sagittarius ($\sim$3 kpc), and the Scutum ($\sim$6 kpc) arm,  but  drops to $\sim$0.7 and $\sim$0.6 in the Outer arm ($\sim$15 kpc), respectively. \cots\, however, decreases significantly with distance, and is approximately zero in the Outer arm.    The sensitivity clip factor shows similar variations with the beam filling factors, but is systematically lower by $\sim$0.2.



\item  The beam filling factor is mainly correlated with the angular size $l$ in the beam size unit  and  can be approximated with $0.922l^2/\left(l+ 0.762\right)^2$. The average beam filling factors of molecular clouds identified with DBSCAN can be derived using this correlation.

\item    We derived a relationship between the observed flux and the mean voxel SNR ($x$), and the ratio of the observed flux to the total flux is approximately $\left(x- 2.224\right)^2/\left(x- 2.224+0.457\right)^2$.  This relationship can be used to estimate the total flux. 


\item The $\eta$-size and $\xi$-sensitivity relationships seem to be universal suggested by the comparison with other existing CO surveys.  
 
\end{enumerate}

\acknowledgments

We would like to show our gratitude to support members of the MWISP group, Xin Zhou, Zhiwei Chen, Shaobo Zhang, Min Wang, Jixian Sun, and Dengrong Lu, and  observation  assistants at PMO Qinghai station for their long-term observation efforts. We are also immensely grateful to other member of the MWISP group, Ye Xu, Hongchi Wang, Zhibo Jiang, Xuepeng Chen, Yiping Ao for their useful discussions. This work was sponsored by the Ministry of Science and Technology (MOST) Grant No. 2017YFA0402701, Key Research Program of Frontier Sciences (CAS) Grant No. QYZDJ-SSW-SLH047,   National Natural Science Foundation of China Grant No. 11773077, U1831136, and 12003071, and the Youth Innovation Promotion Association, CAS (2018355). 

%

\vspace{5mm}
\facilities{PMO 13.7-m}


\software{astropy \citep{2013A&A...558A..33A},     
						SciPy 
          }

 \bibliographystyle{aasjournal}
 \bibliography{refGAIADIS}





%

\end{document}